\documentclass[11pt]{article}

\usepackage{tikz}
\usetikzlibrary{matrix,arrows,shapes}
\usepackage{pgf}
\usepackage{url}
  \usepackage{enumerate}
  \usepackage{amsthm}
  \usepackage{amssymb}
  \usepackage{amsmath}
    \usepackage{amscd}
  \usepackage{eucal}
  \usepackage{mathrsfs}
  \usepackage{upgreek}
  \usepackage{mathtools}
   \usepackage{dsfont}
      \usepackage{yfonts}
  \usepackage[T1]{fontenc}
  \usepackage{bbm}
  \usepackage{geometry}
    \usepackage{graphics}
\usepackage{array}
\usepackage{booktabs}
  \usepackage{enumerate}
 \usepackage{feynmf}
  \usetikzlibrary{trees}
\usetikzlibrary{decorations.pathmorphing}
\usetikzlibrary{decorations.markings}
\tikzset{
 photon/.style={decorate, decoration={snake}, draw=red},
    electron/.style={draw=blue, postaction={decorate},
        decoration={markings,mark=at position .55 with {\arrow[draw=blue]{>}}}},
    gluon/.style={decorate, draw=magenta,
        decoration={coil,amplitude=4pt, segment length=5pt}},
    sderiv/.style={postaction={decorate},
        decoration={markings,mark=at position .3 with {\arrow{>}}}},
    tderiv/.style={postaction={decorate},
        decoration={markings,mark=at position .7 with {\arrow{<}}}},
    stderiv/.style={postaction={decorate},
        decoration={markings,mark=at position .7 with {\arrow{<}},mark=at position .3 with {\arrow{>}}}}
}
\usepackage{soul}
\usepackage[colorinlistoftodos,textsize=footnotesize]{todonotes}

\setstcolor{red}
\newcommand{\CE}{\mathfrak{CE}}  
\newcommand{\E}{\mathfrak{E}}

\newcommand{\V}{\mathfrak{V}}

\newcommand{\fA}{\mathfrak{A}}

\newcommand{\fP}{\mathfrak{P}}
\newcommand{\D}{\mathfrak{D}}

\newcommand{\F}{\mathfrak{F}}
\newcommand{\fG}{\mathfrak{G}}

\newcommand{\KT}{\mathfrak{KT}}

\newcommand{\BV}{\mathfrak{BV}}

\newcommand{\frakg}{\mathfrak{g}}

\newcommand{\G}{\mathcal{G}}  
\newcommand{\Lcal}{\mathcal {L}}
\newcommand{\Bcal}{\mathcal {B}}
\newcommand{\Kcal}{\mathcal{K}}  
\newcommand{\Hcal}{\mathcal{H}}  
\newcommand{\Ccal}{\mathcal{C}}
\newcommand{\Dcal}{\mathcal{D}}
\newcommand{\Ecal}{\mathcal{E}} 
\newcommand{\Fcal}{\mathcal{F}} 

\newcommand{\Ncal}{\mathcal{N}}
\newcommand{\Mcal}{\mathcal{M}}
\newcommand{\Ocal}{\mathcal{O}}
\newcommand{\Acal}{\mathcal{A}}
\newcommand{\Scal}{\mathcal{S}}
\newcommand{\Rcal}{\mathcal{R}}
\newcommand{\Tcal}{\mathcal{T}}

\newcommand{\Wcal}{\mathcal{W}}

\newcommand{\Ci}{\mathcal{C}^\infty} 
\newcommand{\Ca}{\mathrm{\mathbf{C}}}
\newcommand{\Da}{\mathrm{\mathbf{D}}}
\newcommand{\obj}{\mathrm{Obj}}

\newcommand{\Po}{\mathrm{\mathbf{Poi}}}  
\newcommand{\Loc}{\mathrm{\mathbf{Loc}}}       
\newcommand{\Obs}{\mathrm{\mathbf{Obs}}}       
\newcommand{\Vect}{\mathrm{\mathbf{Vec}}}       

\newcommand{\WF}{\mathrm{WF}}         
\newcommand{\id}{\mathrm{id}}               
\newcommand{\relsupp}{{\mathrm{rel}\,\mathrm{supp}}}      
\DeclareMathOperator{\supp}{\mathrm{supp}}      
\newcommand{\Diff}{\mathrm{Diff}}        
\newcommand{\tr}{\mathrm{tr}}                 

\newcommand{\loc}{\mathrm{loc}}
\newcommand{\inv}{\mathrm{inv}}
\newcommand{\reg}{\mathrm{reg}}

\newcommand{\pg}{\mathrm{pg}}

\newcommand{\af}{\mathrm{af}}

\newcommand{\gh}{\mathrm{gh}}

\newcommand{\mc}{{\mu\mathrm{c}}}
\newcommand{\ml}{\mathrm{ml}}
\newcommand{\ex}{\mathrm{ext}}

\newcommand{\dS}{\mathrm{d}\mathbb{S}}
\newcommand{\x}{{\bf x}}
\newcommand{\NN}{\mathbb{N}}          
\newcommand{\RR}{\mathbb{R}}           
\newcommand{\CC}{\mathbb{C}}           
\newcommand{\M}{\mathbb{M}} 	     
\newcommand{\al}{\alpha}

\newcommand{\De}{\Delta}

\newcommand{\la}{\lambda}
\newcommand{\La}{\Lambda}

\newcommand{\ph}{\varphi}
\newcommand{\om}{\omega}
\newcommand{\T}{\cdot_{{}^\Tcal}}

\newcommand{\TT}{\Tcal}

\newcommand{\Poi}[2]{\{#1,#2\}}

\newcommand{\eom}{{\textsc{eom}}}
\newcommand{\qme}{{\textsc{qme}}}

\newcommand{\sst}[1]{\scriptscriptstyle{#1}}  
\newcommand{\minus}{\sst{-1}}   
\newcommand{\pa}{\partial}                              
\newcommand{\be}{\begin{equation}}
\newcommand{\ee}{\end{equation}}

\newcommand{\Lap}{\bigtriangleup}

\global\long\def\bld#1{\boldsymbol{#1}}

\begin{document}
\title{Quantum field theory on curved spacetimes:\\ axiomatic framework and examples}
\author{Klaus Fredenhagen, Kasia Rejzner}
\maketitle
 \theoremstyle{plain}
  \newtheorem{df}{Definition}[section]
  \newtheorem{thm}[df]{Theorem}
  \newtheorem{prop}[df]{Proposition}
  \newtheorem{cor}[df]{Corollary}
  \newtheorem{lemma}[df]{Lemma}
    \newtheorem{exa}[df]{Example}

  \theoremstyle{plain}
  \newtheorem*{Main}{Main Theorem}
  \newtheorem*{MainT}{Main Technical Theorem}

  \theoremstyle{definition}
  \newtheorem{rem}[df]{Remark}

 \theoremstyle{definition}
  \newtheorem{ass}{\underline{\textit{Assumption}}}[section]

\tableofcontents
\markboth{Contents}{Contents}
 \section{Introduction} 
 Since about a century, the relation between quantum physics and gravitation is not fully understood.
Quantum theory is very successful in nonrelativistic physics where precise mathematical results can be compared with experiments,
somewhat less successful in elementary particle physics where theory delivers only the first terms of a formal power series, but up to now it also yields excellent agreement with experiments. General relativity, as the widely accepted theory of gravity, is excellently confirmed by astronomical data and 
deviations can be explained by plausible assumptions (dark matter, dark energy). However, finding a consistent theory which combines both general relativity and quantum physics is still an open problem. 

First steps for investigating the relation between these two fundamental aspects of nature are experiments with slow neutrons in the gravitational field of the earth. This can be treated as a problem in quantum mechanics with the Newtonian gravitational potential, and the experiments are in very good agreement with the theory. A more ambitious problem are the fluctuations of the cosmological microwave background which are explained by quantum fluctuations of the 
gravitational field in the inflationary era. On the theoretical side, the last decades were dominated by attempts to unify general relativity with quantum theory by rather radical new concepts, the best known being string theory and loop quantum gravity. In this report, however, we will concentrate on a much more modest goal to provide a consistent interpretation of existing experimental data in the situation when gravitational forces are weak. In such circumstances one can neglect the backreaction of quantum fields on the gravitational field.

In this overview article we present a formalism suitable for constructing models of QFT's on curved spacetimes. The leading principle is the emphasis on local properties. It turns out that this requires a reformulation of the standard QFT framework which also yields a new perspective for the theories on Minkowski space. The aim of the present work is to provide an introduction  into the framework, which should be accessible for both mathematical physicists and mathematicians. Other recent reviews, which take a different angle include \cite{BDH,HWRev,FewVerchReview}. In particular they discuss some important results not covered here, such as quantum energy inequalities \cite{FewPfe1,Chris2,FewPfe}, Reeh-Schlieder property \cite{Ko09}, PCT and Spin-Statistic theorem \cite{HollandsPCT,Verch01,FewsterReg}. For applications in the study of the back-reaction problem and cosmology see for example \cite{Hack,Hack14,DMP09,DMP09a,DMP11} and a recent book \cite{ThomasBook}.
\section{Lorentzian geometry}

Before we formulate the axioms of quantum field theory on curved spacetimes, we need to introduce some basic geometrical notions used in special and general relativity. Recall that in special relativity space and time are described together with one object, called \textit{Minkowski spacetime}. In general relativity one extends this notion to a certain class of manifolds. We will use the following definition:
\begin{df}
A 4 dimensional spacetime $\Mcal\doteq(M,g)$ is a smooth 4 dimensional manifold (we assume it to be Hausdorff, paracompact, connected) with a smooth pseudo-Riemannian metric $g\in\Gamma(T^*M\otimes T^*M)$, i.e. for every $p\in M$, $g_p$ is a symmetric non degenerate bilinear form of Lorentz signature (we choose the convention $(+,-,-,-)$).
\end{df}
The metric $g$ introduces the notion of \textit{causality}. It is instructive to think of the causal structure as a way to distinguish certain classes of smooth curves, which we call: \textit{timelike}, \textit{null}, \textit{causal} and \textit{spacelike}.
\begin{df}
Let $\gamma: \RR\supset I\rightarrow M$ be a smooth curve in $M$such that $\dot{\gamma}(t)\neq0$ $\forall t\in I$. We say that $\gamma$ is 
\begin{itemize}
\item causal (timelike), if  $g(\dot{\gamma},\dot{\gamma})\geq 0$ ($>0$),
\item spacelike, if  $g(\dot{\gamma},\dot{\gamma})< 0$,
\item null, if  $g(\dot{\gamma},\dot{\gamma})= 0$,
\end{itemize}
where $\dot{\gamma}$ is the vector tangent to the curve.
\end{df}

A spacetime $\Mcal$ is called orientable if there exists a differential form $\omega$ of maximal degree, which does not vanish anywhere  (a volume form). The spacetimes which we considered are always assumed to posses such a volume form. We also assume that our spacetimes are time orientable, i.e. there exists a smooth vector field $v$ which is everywhere timelike. The choice of such a vector field  induces a time orientation.

A timelike or lightlike tangent vector $\xi$ at some spacetime point $x$ is called future directed if 
\[g_x(\xi,v(x))>0\ .\] 
A causal curve is called future directed if all its tangent vectors are future directed. This allows to introduce the future $J_+(x)$ of a point $x$ as the set of points which can be reached from $x$ by a future directed causal curve:
\[J_{+}(x)=\{y\in M| \exists \gamma:x\to y \text{ future directed}\}\] 
In an analogous way, one defines the past $J_-$ of a point. The future (past) of a subset $B\subset M$ is defined by
\[J_\pm(B)=\bigcup_{x\in B}J_\pm(x)\ .\]
Note that future and past are in general not closed. 
Two subsets $O_1$ and $O_2$ in $M$ are called
spacelike separated if they cannot be connected by a causal curve, i.e.\
if for all $x \in \overline{O_1}$,
 $J^{\pm}(x)$ has empty intersection with $\overline{O_2}$.
By $O^{\perp}$ we denote the causal complement of $O$, i.e.\ the
largest open set in $M$ which is spacelike separated from $O$.

Crucial for the following is the concept of global hyperbolicity.

{\begin{df}[after \cite{BeSa}] A spacetime is called globally hyperbolic if it does not contain closed causal curves and if for any two points $x$ and $y$ the set $J_+(x)\cap J_-(y)$ is compact.
\end{df}}

{Globally hyperbolic spacetimes have many nice properties: }
\begin{itemize}

{\item They have a Cauchy surface, i.e. a smooth spacelike hypersurface which is hit exactly once by each nonextendible causal curve.}

{\item They have even a foliation by Cauchy surfaces, and all Cauchy surfaces are diffeomorphic, i.e. globally hyperbolic spacetimes are diffeomorphic to $\Sigma\times\RR$ with Cauchy surfaces $\Sigma\times\{t\},t\in\RR$.}

\item  Normally hyperbolic linear partial differential equations, meaning those with principal symbol $\sigma$ of the metric type (i.e. $\sigma(k)=-g_{\mu\nu}k^\mu k^\nu$) have a well posed Cauchy problem. In particular they have unique retarded and advanced Green's functions (see the beginning of section \ref{freescalar} for precise definitions of these objects). We recall that the principal symbol of a differential operator $P=\sum_{|\al|\leq k}a_\al(x)D^\al$ is defined as $\sigma_P(k)=\sum_{|\al|= k}a_\al(x)k^\al$, where $\al\in \mathbb{N}^n_0$ is a multiindex, i.e. $\alpha = (\alpha_1, \ldots,\alpha_n)$, $\al_i\in\NN$, $|\al|=\al_1+\ldots\al_n$  and $D^\al=\partial_{x_1}^{\alpha_1} \ldots \partial_{x_n}^{\alpha_n}$.
\end{itemize}

Examples of spacetimes:
\begin{enumerate}
\item 
The standard example is \textit{Minkowski space}
\[\mathbb{M}=(\RR^4,\eta)\]
with the metric
\[\eta=dx^0dx^0-dx^1dx^1-dx^2dx^2-dx^3dx^3 \ .\]
Minkowski space is globally hyperbolic, and all spacelike hyperplanes are Cauchy surfaces.
\item
Another example is de Sitter space which can be realized as the hyperboloid
\[\mathrm d\mathbb{S}=\{x\in\mathbb R^5, \eta(x,x)=-1\}\]
in 5 dimensional Minkowski space. Also \textit{de Sitter space} is globally hyperbolic, and intersections of spacelike hyperplanes with the hyperoloid are Cauchy surfaces.
A convenient coordinatization is
\[(t,\Omega):\dS\to \RR\times\mathbb{S}^3\] 
with $t(x)=x^0$ and 
\[\Omega(x)=\frac{\x}{|\x|} \ .\]
The induced metric from the Minkowski space assumes the form
\[g=(1+t^2)^{-1}dt^2-(1+t^2)d\Omega^2\ .\]
with the standard metric $d\Omega^2$ on the 3-sphere $\mathbb{S}^3$.
\item An important example is the \textit{Schwarzschild spacetime} which describes the gravitational field of a non rotating point mass.
It can be described as the set
\[(2m,\infty)\times\RR\times \mathbb{S}^2\]
with the metric
\[\left(1-\frac{2m}{r}\right)dt^2-\left(1-\frac{2m}{r}\right)^{-1}dr^2-r^2d\Omega^2\]
with the standard metric $d\Omega^2$ on the 2-sphere $\mathbb{S}^2$ and $(r,t)\in(2m,\infty)\times\RR$.
\item Schwarzschild spacetime is a special case of a \textit{static spacetime}
\[M=\mathbb{R}\times \Sigma\]
with metric
\[g=a^2dt^2-h\]
where $a$ is a smooth positive function on $\Sigma$ and $h$ is a Riemannian metric on $\Sigma$. Static spacetimes are globally hyperbolic if $\Sigma$ equipped with the so called radiative metric
\[\hat{h}=\frac{1}{a^2}h\]
is a complete Riemannian space. In the case of Schwarzschild spacetime we observe that the boundary at $r=2m$ (Schwarzschild horizon) is at finite distance with respect to the metric $h$,
\[d_h(2m,r)=\int_{2m}^r \left(1-\frac{2m}{r'}\right)^{-\frac12}dr'<\infty\ .\]
{As a consequence, the horizon can be reached in finite proper time.}
But with respect to the radiative metric $\hat h$, the distance is infinite:
\[d_{\hat h}(2m,r)=\int_{2m}^r\left(1-\frac{2m}{r'}\right)^{-1}dr'=\infty\]
{Thus no causal curve reaches the horizon in finite Schwarzschild time $t$. This implies that Schwarzschild spacetime is globally hyperbolic.}

\item \textit{Cosmological (Friedmann-Lemaitre-Robertson-Walker) spacetimes}: these spacetimes are models for the structure of the universe at large scales. They are characterized by a distinguished timelike vector field which might be identified with the tangent vectors of the world lines of galaxy clusters. Moreover, the spatial sections orthogonal to this vector field are maximally symmetric Riemannian spaces, 
i.e. their symmetry group is a 6 dimensional Lie group. This corresponds to the observation of spatial isotropy and homogeneity and is often referred to as the Kopernican principle. Cosmological spacetimes  are of the form
\[M=I\times \Sigma\]
where $I$ is an open interval and $\Sigma$ either the 3 sphere, euclidean 3 space or a unit hyperboloid in 4 dimensional Minkowski space. 
The metric is 
\[g=dt^2-a(t)^2h\]
where $h$ is the standard metric on $\Sigma$,
and $a$ is a smooth positive function on $I$.  Cosmological spacetimes are globally hyperbolic.
\end{enumerate}
\section{Surprises in QFT on curved backgrounds}
Classical field theory, e.g. Maxwell's theory of the electromagnetic field, can easily be formulated on generic Lorentzian manifolds. The electromagnetic field strength is considered as a 2-form
\[
F=F_{\mu\nu}dx^{\mu}\wedge dx^{\nu}
\]
and Maxwell's equation take the form
\[dF=0\]
with the exterior differentiation $d$ and
\[\delta F=j\]
where $\delta=\star^{-1}d\star$ is the codifferential, and $j$ is a conserved current. Let us now review the mathematical structures relevant in this example. First, it is important that spacetime is a smooth orientable manifold $M$ and hence differential forms and exterior derivative are well defined. Another structure used here is a nondegenerate metric $g$ and an orientation. These structures allow to define the Hodge dual as

\[\star:\Omega^k(M)\to\Omega^{n-k}(M),\ n=\mathrm{dim}M \]

\[\star F(X_1,\dots,X_{n-k})d\mu_g=F\wedge g^{\#}(X_1)\wedge\cdots g^{\#}(X_{n-k})\] where $\Omega^k(M)$ is the space of $k$-forms on $M$, $d\mu_g$ is the volume form induced by $g$,
\[d\mu_g(X_1,\dots,X_n)=|\mathrm{det}\left(g(X_i,X_j)_{i,j=1,\dots,n}\right)|^{\frac12}\ ,\]
 $X_1,\dots,X_{n}$ is a positively oriented $n$-tuple of vector fields and the ``musical'' map $g^{\#}:\Gamma(TM)\to \Omega^1(M)$   is defined by
\[g^{\#}(X)(Y)=g(X,Y) \ .\]

We have seen that the essential structures in classical theory can be relatively easily generalized from the Minkowski spacetime to more general Lorentzian manifolds. The standard formalism of QFT, however, relies heavily on Poincar\'{e} symmetry, which is characteristic for Minkowski space. Note that in standard QFT:
\begin{itemize}

{\item Particles are defined as irreducible representations of the (covering of the) Poincar\'{e} group.}

{\item There is a distinguished state, the vacuum, understood as the state with all particles absent.}

{\item The main physical observable is the S-matrix, describing the transition from incoming to outgoing particle configurations.}

{\item  Momentum space (as the dual of the subgroup of translations) plays an important role for calculations.}

{\item Transition to imaginary time (euclidean QFT) is often helpful in order to improve convergence. }
\end{itemize}


None of these features is present for QFT on generic Lorentzian spacetimes. First of all, the group of spacetime symmetries of a generic spacetime is trivial. Accordingly, the very concept of particles is no longer available and the idea of the vacuum as the state with no particles becomes meaningless. Also,  transition to imaginary times (and a corresponding transition to a Riemannian space) is possible only in special cases. Moreover, the Fourier transform is in general not defined, so calculations relying on momentum space cannot be done. Finally, there is no unique definition of the Feynman propagator. All these facts lead to  some peculiarities of quantum field theory on generic spacetimes, which include:
\begin{itemize}

{\item ``Particle creation'':} 

{In free field theory, one might introduce a particle concept which is appropriate for some spatial hypersurface.} 
{But comparison of the particle numbers on different hypersurfaces yields particle creation
(typically infinite changes).}

{\item ``Hawking radiation'':} 

{In the analysis of a scalar field in the field of a collapsing star one finds that an initial ground state (in the static situation before the collapse) evolves into a state with thermal radiation after the collapse.}

{\item ``Unruh effect'': }

{Even on Minkowski space, for a uniformly accelerated observer, the vacuum gets thermal properties.}
\end{itemize}

The most prominent feature of QFT on curved spacetime is the lack of a distinguished vacuum state. There were numerous attempts to get around this problem. Here we want to list some of them. One of the attempts was the concept of \textit{instantaneous vacuum states} (defined as ground states of the instantaneous Hamiltonian). This, however, turned out to depend in a very singular way on the choice of the spatial surface. \textit{Adiabatic vacuum states}, originally introduced by  Parker \cite{Parker} and based on a WKB approximation, are  better behaved, but not unique. \textit{States of low energy} (Olbermann \cite{Olbermann}), defined as ground states with respect to a time averaged Hamiltonian, are  well behaved, but depend on the averaging. Another idea was to consider states evolving from a vacuum state in past asymptotically flat spacetimes (\cite{DMP09a,DMP09,DMP11,DHP11,DS11,BDM14}). Finally, there is the Sorkin-Johnston state \cite{Johnston,Sorkin,SJ12}: such a state can be uniquely constructed from the geometrical data, but is  in general too singular (Fewster-Verch \cite{FV12}) for a definition of relevant observables as e.g. the energy momentum tensor. A modified, ``smoothed-out'' version of this construction was recently proposed in \cite{BrumF14}.

Typically, these proposed states can be defined only in special cases (free scalar field, cosmological spacetimes). In view of these problems, we use the following strategy for the formulation of QFT on curved spacetime:
\begin{enumerate}
\item Decouple local features (field equations, commutation relations) from nonlocal features (correlations). This amounts to construct, in the first step, the algebra of observables as an abstract algebra and consider afterwards representations by Hilbert space operators (``algebraic approach to QFT''). On Minkowski space this approach is summarized by the Haag-Kastler axioms \cite{HK,Haag}.
\item Find a local version of the spectrum condition (``positivity of energy'') which is the most important structural impact of the Hilbert space representation of QFT on Minkowski space. This can be done with the use of the notion of the class of Hadamard states. The idea that such a class, rather than a single vacuum state, is the crucial structure in QFT on curved spacetime was proposed by Kay \cite{Kay84} and Haag-Narnhofer-Stein  \cite{HNS}.
\end{enumerate}
\section{Algebraic approach to QFT on curved spacetimes}
\subsection{Haag-Kastler axioms}\label{AQFTaxioms}
One of the motivations for developing quantum field theory (on Minkowski space) was to formulate a framework which allows to incorporate the basic principles of both quantum mechanics and special relativity. At first sight it might seem that these two theories operate with completely different and maybe even contradictory paradigms. In quantum mechanics, which is the theory of the universe in small scales, one works with operators on Hilbert spaces and interprets the theory using some notions from probability and statistics. An important feature of quantum mechanics is the existence of correlations and entanglement. In special relativity, on the other hand, the basic principle is that of causality, which states that observations performed in spacelike separated regions cannot influence each other. In order to see how entanglement can coexist with causality, we need to distinguish between the properties of observables and the properties of states. We will see in this section that in algebraic quantum field theory the observables are local, whereas the states can exhibit long range correlations.

The basic principle underlying the algebraic approach to QFT is \textit{locality}. In this context it is realized by identifying the algebras of observables that can be measured in given bounded regions of spacetime. In other words we associate to each bounded $\Ocal\subset \M$ a $C^*$-algebra $\fA(\Ocal)$.  Note that the collection $\mathcal{B}$ of bounded subsets of $\M$ is a directed set, i.e. it is equipped with a reflexive and transitive binary relation (in this case $\subseteq$, such that for each pair $\Ocal_1,\Ocal_2\in\mathcal{B}$, there exists $\Ocal\in\mathcal{B}$, such that $\Ocal_1\subseteq \Ocal$ and $\Ocal_2\subseteq \Ocal$. A function from a directed set to some other set is called \textit{a net}, so the assignment which we have just defined is \textit{a net of $C^*$-algebras}. Now let us come back to some physical considerations, which imply further properties of the net $\{\fA(\Ocal)\}_{\Ocal\in\mathcal{B}}$. First, the association of algebras to spacetime regions has to be compatible with a physical notion of subsystems. It means that if we have a region $\Ocal$ which lies inside $\Ocal'$ we want the corresponding algebra $\fA(\Ocal)$ to be contained inside $\fA(\Ocal')$, i.e. in a bigger region we have more observables. This property can be formulated as the Isotony condition for the net $\{\fA(\Ocal)\}_{\Ocal\in\mathcal{B}}$ of local algebras associated to bounded regions of the spacetime. In the Haag-Kastler framework one imposes some further, physically motivated, properties. We say that a net of algebras satisfies the Haag-Kastler axioms if the following hold:
\begin{itemize}
	\item {\bf Isotony}. For $\Ocal\subset \Ocal'$ we have $\fA(\Ocal)\subset \fA(\Ocal')$.
	\item {\bf Locality (Einstein causality)}. Algebras associated to spacelike separated regions commute: if
$\Ocal_1$ is spacelike separated from $\Ocal_2$, then $[A, B] = 0$, $\forall A \in \fA(\Ocal_1)$, $B \in \fA(\Ocal_2)$. This expresses the ``independence'' of physical systems associated to regions  $\Ocal_1$ and $\Ocal_2$. The \textit{quasilocal algebra} is defined as the inductive limit $\fA(\M)\doteq\overline{\bigcup\limits_\Ocal\fA(\Ocal)}$, and for any (possibly unbounded) region $\Ncal\subset\M$ the algebra $\fA(\Ncal)$ is the C*-subalgebra of $\fA(\M)$ generated by the algebras $\fA(\Ocal)$ with $\Ocal\subset \Ncal$.
	\item {\bf Covariance}. The Minkowski spacetime has a large group of isometries, namely the connected component of the Poincar\'e group. We require that  there exists a family of isomorphisms $\alpha_L^\Ocal : \fA(\Ocal) \rightarrow\fA(L \Ocal)$ for Poincar\'e transformations $L$, such that for
$\Ocal_1 \subset \Ocal_2$ the restriction of $\alpha_L^{\Ocal_2}$ to $\fA(\Ocal_1)$ coincides with $\alpha_L^{\Ocal_1}$ and such that $\alpha^{L\Ocal}_{L'}\circ \alpha^{\Ocal}_{L}=\alpha^{\Ocal}_{L'L}$.
	\item {\bf Time slice axiom}: The solvability of the initial value problem can be formulated as the requirement that the algebra $\mathfrak{A}(\Ncal)$ of any causally convex neighborhood $\Ncal$ of some Cauchy surface $\Sigma$ already coincides with $\mathfrak{A}(\M)$. This means in particular that we only need to determine our observables in some small time interval $(t_0-\epsilon,t_0+\epsilon)$ to reconstruct the full algebra.
	\item {\bf Spectrum condition}. This condition corresponds physically to the positivity of energy. One assumes that there exists a compatible family of faithful representations $\pi_{\Ocal}$ of 
	$\fA(\Ocal)$ on a fixed Hilbert space (i.e. the restriction of $\pi_{\Ocal_2}$ to $\fA(\Ocal_1)$ coincides with $\pi_{\Ocal_1}$ for $\Ocal_1\subset\Ocal_2$) such that translations are unitarily implemented, i.e. there is a unitary representation $U$ of the translation group satisfying
	\[U(a)\pi_{\Ocal}(A)U(a)^{-1}=\pi_{\Ocal+a}(\alpha_a(A))\ ,\ A\in\fA(\Ocal),\]
	and such that the joint spectrum of the generators $P_{\mu}$ of translations $e^{iaP}=U(a)$, $aP=a^\mu P_\mu$, is contained in the closed forward lightcone: $\sigma(P)\subset \overline{V}_+$. 
	\end{itemize}
\begin{rem}
For this definition of \textit{Covariance}, one doesn't need to assume that $\mathcal{B}$ is a directed set ($\{\fA(\Ocal)\}_{\Ocal\in\mathcal{B}}$ doesn't need to be a net), hence it is applicable for spacetimes $\Mcal$ much more general than Minkowski if we replace the connected component of Poincar\'e group with the isometry group of $\Mcal$. If $\{\fA(\Ocal)\}_{\Ocal\in\mathcal{B}}$ is indeed a net, one can take an inductive limit and the family of isomorphisms $\alpha_L^\Ocal$ induces a representation $\al$ by automorphisms $\al_L$, such that $\al_L(\fA(\Ocal))=\fA(L\Ocal)$. 
\end{rem}
\subsection{Axioms of Locally Covariant Quantum Field Theory}\label{axioms}
We now want to generalize the framework introduced in section \ref{AQFTaxioms} to quantum theories on generic spacetimes. To start with, we may think of a globally hyperbolic neighborhood  $U$ of a spacetime point $x$ in some spacetime  $\Mcal=(M,g)$. 
Moreover, we assume that any causal curve in $M$ with end points in $U$ lies entirely 
in $U$. Then we require that the structure of the algebra of observables $\fA(U)$ associated to $U$ should be completely independent of the world outside. Let us formalize this idea. We call an embedding $\chi:\Mcal\to \Ncal$ of a globally hyperbolic spacetime $\mathcal{M}$ into another one $\Ncal$ \textit{admissible} if it is an isometry, it preserves the metric, orientations, the causal structure. The property of \textit{causality preserving} is defined as follows: let $\chi:\Mcal\to \Ncal$, for any causal curve $\gamma : [a,b]\to N$, if $\gamma(a),\gamma(b)\in\chi(M)$ then for all $t	\in ]a,b[$ we have: $\gamma(t)\in\chi(M)$ . We also assume that $\chi$ is such that the source manifold is topologically ``simpler'' than the target. The precise definition depends on the concrete theory. In the examples studied in this review it is sufficient to assume that the source manifold has $H^1(M)=H^2(M)={0}$. Different theories can be sensitive to different features of a spacetime. The literature concerning these topological aspects will be briefly reviewed in section \ref{topo}. 

We require that for each such admissible embedding there exists an injective homomorphism 
\be
\alpha_{\chi}:\mathfrak{A}(\Mcal)\to\mathfrak{A}(\Ncal)
\ee
of the corresponding algebras of observables assigned to them, moreover if $\chi_1:\Mcal\to \Ncal$ and $\chi_2:\Ncal\to \Lcal$ are embeddings as above then we require the covariance relation
\be\label{cov}
\alpha_{\chi_2\circ\chi_1}=\alpha_{\chi_2}\circ\alpha_{\chi_1} \ .
\ee
Such an assignment $\mathfrak{A}$ of algebras to spacetimes and algebra-morphisms to embeddings can be interpreted in the language of category theory as a \textit{covariant functor} between two categories: the category $\Loc$ of globally hyperbolic, oriented and time-oriented spacetimes with admissible embeddings as arrows and the category $\Obs$ of unital $C^*$-algebras. Intuitively, a category consists of a class of objects and maps between them, called arrows, satisfying certain axioms. For more details on categories and functors, see appendix \ref{categories}. The requirement that $\fA$ is a covariant functor  already generalizes the Haag-Kastler axioms of \textit{Isotony} and \textit{Covariance} (equation \eqref{cov}). We say that a functor $\mathfrak{A}: \Loc\rightarrow \Obs$ is a \textit{locally covariant quantum field theory} (LCQFT), if it satisfies two further properties:
\begin{itemize}
\item \textbf{Einstein causality}: let $\chi_i:\Mcal_i\rightarrow \Mcal$, $i=1,2$ be morphisms of $\Loc$ such that $\chi_1(M_1)$ is causally disjoint from $\chi_2(M_2)$, then we require that:
\[
[\al_{\chi_1}(\fA(\Mcal_1)),\al_{\chi_2}(\fA(\Mcal_2))]=\{0\}\,,
\]
\item \textbf{Time-slice axiom}: let $\chi:\Ncal\rightarrow \Mcal$, if $\chi(\Ncal)$ contains a neighborhood of a Cauchy surface $\Sigma\subset M$, then $\al_\chi$ is an isomorphism.
\end{itemize}

The \textbf{Einstein causality} requirement reflects the commutativity of observables localized in spacelike separated regions. From the point of view of category theory, this property is encoded in the tensor structure of the functor $\mathfrak{A}$. In order to make this statement precise, we need to equip our categories $\Loc$ and $\Obs$ with tensor structures (for a precise definition of a tensor category, see \ref{categories}).

The category of globally hyperbolic manifolds $\Loc$ can be extended to a monoidal category $\Loc^\otimes$, if we extend the class of objects with finite disjoint unions of elements of $\obj(\Loc)$, 
\[\Mcal=\Mcal_1\sqcup \ldots\sqcup \Mcal_k\,,
\] 
where $\Mcal_i\in\obj(\Loc)$. Morphisms of $\Loc^\otimes$ are isometric embeddings, preserving orientations and causality. More precisely, they are maps $\chi: \Mcal_1\sqcup \ldots\sqcup \Mcal_k\rightarrow  \Mcal$ such that each component satisfies the requirements for a morphism of $\Loc$ and additionally all images are spacelike to each other, i.e., $\chi(M_i) \perp\chi(M_j)$, for $i\neq j$. $\Loc^\otimes$ has the disjoint union as a tensor product, and the empty set as unit object. It is a monoidal category and, using the results of \cite{JoSt}, it is tensor equivalent to a strict monoidal category, which we will denote by the same symbol $\Loc^\otimes$.

On the level of C*-algebras the choice of a tensor structure is less obvious, since, in general, the algebraic tensor product $\fA_1\odot\fA_2$ of two $C^*$-algebras can be completed to a $C^*$-algebra with respect to many non-equivalent tensor norms. The choice of an appropriate norm has to be based on some further physical indications. This problem was discussed in \cite{BFIR}, where it is shown that  
a physically justified tensor norm is the minimal $C^*$-norm $\|.\|_{\textrm{min}}$ defined by
\[
\|A\|_{\textrm{min}} \doteq \sup\{\|(\pi_1\otimes\pi_2)(A)\|_{\Bcal(\Hcal_1\otimes\Hcal_2)}\}\ , \quad A\in\fA_1\otimes\fA_2\,,
\]
where $\pi_1$ and $\pi_2$ run through all representations of $\fA_1$ and of $\fA_2$ on Hilbert spaces $\Hcal_1$, $\Hcal_2$ respectively. $\Bcal$ denotes the algebra of bounded operators. If we choose $\pi_1$ and $\pi_2$ to be faithful, then the
supremum is achieved, i.e. $\|A\|_{\textrm{min}}=\|(\pi_1\otimes\pi_2)(A)\|_{\Bcal(\Hcal_1\otimes\Hcal_2)}$.
The completion of the algebraic tensor product $\fA_1\odot\fA_2$  with respect to the minimal norm $\|A\|_{\textrm{min}}$ is denoted by $\fA_1\underset{\sst{\textrm{min}}}{\otimes}\fA_2$. It was shown in \cite{BFIR} that, under some technical assumptions, 
a functor $\fA:\Loc\rightarrow\Obs$ satisfies the axiom of Einstein causality if and only if it can be extended to a tensor functor $\fA^\otimes:\Loc^\otimes\rightarrow\Obs^\otimes$, which means that
\begin{eqnarray}
\fA^\otimes\left(\Mcal_1\sqcup \Mcal_2\right)&=&\fA^\otimes(\Mcal_1)\otimes_{\textrm{min}}\fA^\otimes(\Mcal_2)\,,\\
\fA^\otimes(\chi\otimes\chi')&=&\fA^\otimes(\chi)\otimes\fA^\otimes(\chi')\,,\\
\fA^\otimes(\varnothing)&=&\CC\,.
\end{eqnarray}

Let us now turn to the \textbf{time slice axiom}. We can use it to describe the evolution between different Cauchy surfaces. As a first step we associate to each Cauchy surface $\Sigma$ the inverse limit
\be\label{alg:Sigma}
\mathfrak{A}(\Sigma)=\lim^{\leftarrow}_{\Ncal\supset \Sigma}\mathfrak{A}(\Ncal)\ .
\ee
Let  us denote by $\iota_{\Mcal\Ncal}$ the inclusion of a causally convex subset $\Ncal$ in $\Mcal$ and by $\al_{\Mcal\Ncal}\doteq\fA\iota_{\Mcal\Ncal}$, the corresponding morphism in $\hom(\Obs)$. Elements of the inverse limit \eqref{alg:Sigma} are sequences $A=(A_\Ncal)_{\Lcal_A\supset \Ncal\supset \Sigma}$ with $\alpha_{\Kcal\Ncal}(A_\Ncal)=A_\Kcal$, $\Kcal\subset \Lcal_A$, with the equivalence relation
\be
A\sim B \text{ if }\exists\ \Lcal\subset \Lcal_A\cap \Lcal_{B}\text{ such that }A_\Ncal=B_\Ncal \text{ for all }\Ncal\subset \Lcal\ . 
\ee
The algebra $\mathfrak{A}(\Sigma)$ can be embedded into $\mathfrak{A}(\Mcal)$ by
\be\label{MSig}
\alpha_{\Mcal\Sigma}(A)=\alpha_{\Mcal\Ncal}(A_\Ncal) \text{ for some (and hence all) }\Sigma\subset \Ncal\subset \Lcal_A\ .
\ee
The concrete realization of $\al_{\Mcal\Sigma}$ for the example of the scalar field is given at the end of section \ref{freescalar}. If we now adopt the time slice axiom, we find that each homomorphism $\alpha_{\Mcal\Ncal}$ is an isomorphism. Hence $\alpha_{\Mcal\Sigma}$ is also an isomorphism, and we obtain the propagator between two Cauchy surfaces $\Sigma_1$ and $\Sigma_2$ by
\be
\alpha^\Mcal_{\Sigma_1\Sigma_2}=\alpha_{\Mcal\Sigma_1}^{-1}\circ\alpha_{\Mcal\Sigma_2}\,.
\ee
This construction resembles constructions in topological field theory for the description of cobordisms, but in the latter case, one associates (finite dimensional) Hilbert spaces to components of the boundary and maps between these Hilbert spaces to the spacetime itself. In the present formulation one avoids the problems related to generalizing the topological field theory construction to the case of infinite dimensional Hilbert spaces. This is because we don't work on the level of concrete Hilbert spaces, but rather on the level of abstract $C^*$-algebras. 

In the Haag-Kastler framework on Minkowski space an essential ingredient was translation symmetry. This symmetry allowed the comparison of observables in different regions of spacetime and was  (besides locality) a crucial input for the analysis of scattering states. In the general covariant framework sketched above no comparable structure is available. Instead one may use fields which are subject to a suitable covariance condition, termed locally covariant fields. A locally covariant field is a family $\ph_\Mcal$ of fields on spacetimes $\Mcal$ such that for every embedding $\chi\in\hom(\Loc)$, $\chi:\Mcal\to \Ncal$ as above
\be
\alpha_{\chi}(\ph_\Mcal(x))=\ph_\Ncal(\chi(x)) \ .
\ee
If we consider fields as distributions with values in the algebras of observables, a field $\ph$ may be considered as a natural transformation from the functor $\mathfrak{D}$ of test function spaces to the functor $\mathfrak{A}$ of field theory.
The functor $\mathfrak{D}$ associates to every spacetime $\Mcal$ its space of compactly supported $\mathcal{C}^{\infty}$-functions,
\be
\mathfrak{D}(\Mcal)=\mathcal{C}^{\infty}_{\mathrm{c}}(M,\RR)\ ,
\ee
and to every embedding $\chi:\Mcal\to \Ncal$ of spacetimes the pushforward of test functions $f\in\mathfrak{D}(\Mcal)$
\be
\mathfrak{D}\chi\equiv \chi_*\ ,\ \chi_*f(x)=\left\{ \begin{array}{ccc}
                                                                                    f(\chi^{-1}(x)) &,&x\in\chi(M)\\
                                                                                    0                    &,&\text{ else}
                                                                                    \end{array}
                                                                           \right.\,.
\ee
$\mathfrak{D}$ is a covariant functor. Its target category is the category of locally convex topological vector spaces $\Vect$ which contains also the category of topological algebras which is the target category for $\mathfrak{A}$. A natural transformation $\ph:\mathfrak{D}\to\mathfrak{A}$ between covariant functors with the same source and target categories is a family of morphisms $\ph_{\Mcal}:\mathfrak{D}(\Mcal)\to\mathfrak{A}(\Mcal)$, $\Mcal\in \mathrm{Obj}(\Loc)$ such that
\be
\mathfrak{A}\chi\circ\ph_{\Mcal}=\ph_{\Ncal}\circ\mathfrak{D}\chi
\ee
with $\mathfrak{A}\chi=\alpha_{\chi}$. We denote the space of locally covariant quantum fields by $\Fcal_q$.
\subsection{Example: Weyl algebra of a free scalar field}\label{freescalar}
Let us consider an example of a locally covariant quantum field theory satisfying the axioms introduced above. It was proven by Dimock in \cite{Dim} that the construction of the Weyl algebra of a free scalar field can be performed on generic globally hyperbolic spacetimes. In \cite{BFV} it was proven, that the assignment of such Weyl algebras to spacetimes defines a functor which is a LCQFT. Here we recall briefly this argument. Let $\Mcal=(M,g)\in\obj(\Loc)$ and let $\E(\Mcal)\doteq\Ci(M,\RR)$. The Klein-Gordon operator  is given by
\[
\square_g+m^2\,,
\]
with the d'Alembertian
{\[\square_g=\delta d=|\det{g}|^{-\frac12}\partial_{\mu}(g^{-1})^{\mu\nu}|\det{g}|^{\frac12}\partial_{\nu}\]
in local coordinates and with the mass $m$.  Let $P\doteq -(\Box_g+m^2)$. From the global hyperbolicity and the fact that $P$ is normally hyperbolic follows that the Cauchy problem for the equation 
\[
P\ph=0\,,\quad \ph\in\mathfrak{E}(\mathcal{M})
\]
is well posed and, as shown in \cite{Baer}, there exist retarded and advanced Green's operators, i.e. linear operators  $\Delta^{R/A}:\D(\Mcal)\rightarrow\E(\Mcal)$ (here $\D(\Mcal)\doteq\Ci_c(M,\RR)$ is the space of compactly supported configurations)  uniquely characterized by the conditions
\[P\circ \Delta^{R/A}=\Delta^{R/A}\circ P=\mathrm{id}_{\mathfrak{D}(\mathcal{M})}\]
and the support condition
\[\mathrm{supp} \ \Delta^{R/A}f\subset J_{\pm}(\mathrm{supp} \  f)\]
with the future (past ) $J_{\pm}$ of a subset of spacetime. 
\begin{exa}
We can illustrate the general properties of propagators on globally hyperbolic spactimes with two instructive examples.
\begin{enumerate}
\item
On Minkowski space
\[\mathbb{M}=(\RR^4,\eta)\]
with the metric
\[\eta=dx^0dx^0-dx^1dx^1-dx^2dx^2-dx^3dx^3 \ .\]
the wave equation (massless Klein-Gordon equation) has the propagators
\[(\Delta^{R/A}f)(x)=\frac{1}{4\pi}\int \frac{f(x^0\mp |\mathbf x-\mathbf y|,\mathbf y)}{|\mathbf x-\mathbf y|}d^3\mathbf y \ .\]
\item
On a globally hyperbolic static spacetime 
\[(\RR\times\Sigma, g=a^2(dt^2-\hat{h})),\] 
we have 
\[g^{-1}=\frac{1}{a^2}(\partial_t\otimes\partial_t-\hat{h}^{-1})\] 
and
\[d\mu_g=a^4dt\wedge d\mu_{\hat{h}}\ ,\]
hence the d'Alembertian takes the following form:
\[\square_g=\frac{1}{a}(\partial_t^2-K)\frac{1}{a}\]
with 
\[K=a\left(|\det{g}|^{-\frac12}\partial_{j}(\hat{h}^{-1})^{jk}|\det{g}|^{\frac12}a^{-2}\partial_{k}\right)a\] 
$K$ is a positive selfadjoint operator on the Hilbert space $L^2(\Sigma,\sqrt{|\mathrm{det}g|})$ and is essentially selfadjoint on $\mathcal{C}^{\infty}_c(\Sigma,\mathbb{R})$
(\cite{Chernoff,Kay78,BrumF14}).
{This allows to write the propagators in terms of the spectral resolution of $K$: Let $f_t:=f(t,\cdot)$ for $f\in\mathcal{C}^{\infty}(\Sigma\times\mathbb{R},\mathbb{R})$. Then
\[(\Delta^Rf)_t=\int_{-\infty}^t \frac{\sin\sqrt{K}(t-s)}{\sqrt{K}}f_sds\ .\]}
{This formula can be checked easily by differentiating twice with respect to the time variable:} 
{\[\frac{\partial}{\partial t}(\Delta^Rf)_t=\int_{-\infty}^t \cos{\sqrt{K}(t-s)}f_sds \]}
{\[\frac{\partial^2}{\partial t^2}(\Delta^Rf)_t=f_t-K(\Delta^Rf)_t\ .\]}
\end{enumerate}
\end{exa}
The difference
\[\Delta=\Delta^R-\Delta^A\]
(often called the causal propagator, and later named the commutator function) has the properties:
\begin{enumerate}
\item For every $f\in\D(\Mcal)$, $\Delta f$ is a solution of the Klein-Gordon equation, and $\supp\ \Delta f\cap\Sigma$ is compact for every Cauchy surface $\Sigma$ (we say that $\Delta f$ has {{spacelike compact support}}).
\item Every smooth solution of the Klein-Gordon equation with spacelike compact support is of the form $\Delta f$ for some $f\in \mathcal \D(\Mcal)$.
\item The bilinear form on $\mathcal \D(\Mcal)$
\be\label{bilinear}
\sigma(f,h)=\int f(\Delta h) d\mu_{g}\ ,
\ee
is antisymmetric, and $\sigma(f,h)=0$ for all $h\in\D(\Mcal)$ if and only if $f=Pf'$ for some $f'$ with compact support. Here $d\mu_{g}$ is the metric-induced volume form on $M$.
\end{enumerate}
To prove the second property, let $\ph$ be a solution, $\chi \in \mathcal{C}^\infty(M,\RR)$, and $\Sigma_1, \Sigma_2$ be Cauchy surfaces such that $\Sigma_1 \cap J_+(\Sigma_2)=\emptyset$. Assume $\chi(x)=0$ for $x \in J_-(\Sigma_1)$ and $\chi(x)=1$ for $x \in J_+(\Sigma_2)$. Then $P \chi \ph=0$ outside $\Sigma_1, \Sigma_2$ ($\chi =$const. there) which implies $P\chi \ph$ has compact support and we can set $f=P\chi \ph$. Hence,
\[
 \Delta f= \Delta P \chi \ph =  \Delta^R P \chi \ph +  \Delta^A P(1-\chi)\ph = \ph.
\]

We can come back now to the construction of the Weyl algebra of the free scalar field. We denote the range
$\De\D(\Mcal)$ by $\mathcal{R}$, i.e.
\[
\Rcal=\{\ph \in \mathcal{C}^\infty(M), P\ph=0 \textrm{ s.t. initial data have compact support}\}
\]
The bilinear form \eqref{bilinear} induces a symplectinc form on $\Rcal$ given by
\[
\sigma(\ph_1,\ph_2)=\int f(\Delta h) d\mu_{g}\,,
\]
where $\ph_1\equiv \Delta f$ and $\ph_2\equiv \Delta h$. Equivalently (see \cite{Baer} for details) this can be written as
\[
\sigma(\ph_1,\ph_2) = \int_\Sigma  (\ph_1 (\partial_n \ph_2) - (\partial_n \ph_1)\ph_2)dvol_\Sigma,
\]
where $\partial_n$ is the normal derivative on $\Sigma$ ($\partial_n \ph_1 = n^{\mu} \partial_{\mu}\ph_1$, $n^{\mu} \xi_{\mu}=0 $ for $\xi \in T\Sigma$, $n^{\mu}n_{\mu}=1$). It can
be shown \cite{Dim} that $({\mathcal R},\sigma)$ is
a symplectic space. There are also other equivalent characterizations of $({\mathcal R},\sigma)$ that are closer to the canonical formalism of classical mechanics.

First, we express the symplectic form $\sigma$ in terms of the Cauchy data. Recall that on a globally hyperbolic space-time, each solution of the Klein-Gordon equation is characterized by its Cauchy data on a Cauchy surface $\Sigma$. Let $\iota_{M\Sigma}$ be the embedding of $\Sigma$ in $\Mcal$. The one to one correspondence between solutions and Cauchy data means that for every pair  $(\phi_1, \phi_2)$, where $\phi_1\in\Ci_c(\Sigma,\RR)$ and $\phi_2$ is a compactly supported density on $\Sigma$, there exists the unique $\ph\in\E(\Mcal)$ such that $P\ph=0$ and  $(\phi_1, \phi_2)=(\iota_{\Mcal\Sigma}^*\ph, \iota_{\Mcal\Sigma}^*(*d\ph))$. We denote the space of compactly supported densities (i.e. compactly supported sections of the determinant bundle) by $\Ci_{dc}(\Sigma,\RR)$. This way we obtain the space
\[
L_2 = \{ (\phi_1, \phi_2) \in \mathcal{C}_c^\infty(\Sigma,\RR) \times \mathcal{C}_{dc}^\infty(\Sigma,\RR) \}
\]
with the symplectic form
\[
\sigma_2((\phi_1, \phi_2), (\psi_1, \psi_2)) = \int_\Sigma (\phi_1 \psi_2 - \phi_2 \psi_1).
\] 
This characterisation is more in line with classical mechanics, since $ \mathcal{C}_c^\infty(\Sigma) \times   \mathcal{C}_{dc}^\infty(\Sigma,\RR)$ plays a role of the classical phase space. 
%

Let us come back to the quantization of the free scalar field. We can associate to $(\Rcal,\sigma)$ its Weyl-algebra $\mathfrak{W}({\mathcal R},\sigma)$, which is generated
by a family of unitary elements $\Wcal(f)$, $\De f \in {\mathcal R}$,
satisfying the Weyl relations,
\be\label{Weyl}
\Wcal(f)\Wcal(h) = {\rm
  e}^{-i\sigma(f,h)/2} \Wcal(f + h)\,.
  \ee
Setting
 $\fA(M,g) = \mathfrak{W}({\mathcal
  R}(M,g),\sigma_{(M,g)})$,
we obtain a covariant functor satisfying the LCQFT axioms. Details of the proof can be found in \cite{BFV}. We can now give an explicit formula for the map $\al_{M\Sigma}$ defined abstractly by \eqref{MSig}. The Weyl algebra associated to a Cauchy surface $\Sigma$ is given by the Weyl quantization of  $(L_2,\sigma_2)$, with generators $\widetilde{\Wcal}(f_1,f_2)$, so we have
\[
\al_{M\Sigma}(\widetilde{\Wcal}(f_1,f_2))=\Wcal(f)\,,
\]
where $\De f$ is the solution with the Cauchy data $(f_1,f_2)$.
\subsection{Example: recovering the Haag-Kastler axioms}
To end this section we will show that using the general framework of LCQFT one recovers the Haag-Kastler axioms for a net of algebras on a fixed spacetime $\Mcal$.
To do this, we restrict the category of spacetimes to subregions of a given spacetime and the arrows to inclusions. In this way we obtain the Haag-Kastler net of local algebras on a globally hyperbolic spacetime as introduced by Dimock \cite{Dim}. In case the spacetime has nontrivial  isometries,  we obtain additional embeddings, and the covariance condition above provides a representation of the group of orientation preserving isometries by automorphisms of the Haag-Kastler net (precosheaf of algebras). 

Let $\Kcal(\Mcal)$ be the family of relatively compact, causally convex subsets of $\Mcal$.  We have $\mathcal{K}(\mathcal{M})\ni\Ocal=(O,g_{O})$, where $O\subset M$ and $g_{O}$ is the restriction of the metric $g$ to $O$. 
 We define a net of algebras $\Ocal\mapsto \Acal(\Ocal)$ by setting: $\Acal(\Ocal)\doteq \al_{\Mcal\Ocal}(\fA(\Ocal,g_{\Ocal}))$, where $\al_{\Mcal\Ocal}\doteq \mathfrak{A}\iota_{\Mcal\mathcal{O}}$. We denote by $\Acal$ the minimal $C^*$-algebra generated by all $\Acal(\Ocal)$, $\Ocal\in\Kcal(\Mcal)$. Now we prove how the covariance in the sense of Haag-Kastler axioms arises in the framework of LCQFT. 
 
 Let $\mathrm{Iso}(\Mcal)$ be the group of isometries of $\Mcal$. For each  $\kappa\in\mathrm{Iso}(\Mcal)$ we have a $*$-homomorphism $\fA\kappa:\fA(\Mcal)\rightarrow \fA(\Mcal)$. Note that $\kappa$ induces also a map $\tilde{\kappa}:\Ocal\rightarrow \kappa(\Ocal)$, $x\mapsto \kappa(x)$ and hence a $*$-homomorphism $\fA\tilde{\kappa}:\fA(\Ocal)\rightarrow \fA(\kappa(\Ocal))$. A short calculation shows that
\begin{align*}
\fA\kappa(\Acal(\Ocal))&=(\fA\kappa\circ\al_{\Mcal\Ocal})(\fA(\Ocal)=\fA({\kappa\circ\iota_{\Mcal\Ocal}})(\fA(\Ocal))=\\
&=\fA({\iota_{\Mcal\kappa(\Ocal)}\circ\tilde{\kappa}})(\fA(\Ocal))=\al_{\Mcal\kappa(\Ocal)}\circ\fA{\tilde{\kappa}}(\fA(\Ocal))=\\
&=\al_{\Mcal\kappa(\Ocal)}(\fA(\kappa(\Ocal)))=\Acal(\kappa(\Ocal))\,,
\end{align*}
where we used the properties of $\tilde{\kappa}$ and covariance of $\fA$. The above calculation shows that we obtain a representation $\tilde{\al}_\kappa$ of $\mathrm{Iso}(\Mcal)$ by  $*$-homomorphisms of $\Acal$ by setting $\tilde{\al}_\kappa\doteq\fA\kappa\upharpoonright_{\Acal}$.

\section{Constructing models}
Models in QFT and also in LCQFT can be constructed in many different ways. Here we use the approach which is close to our classical intuition, where we start with a classical field theory and then quantize it using a certain deformation of the pointwise product. 
\subsection{Classical theory}
\subsubsection{Generalized Lagrangian formalism}\label{genLagr}
For simplicity, let us first consider a scalar field theory. On a given spacetime $\mathcal{M}=(M,g)$ the possible field configurations are the smooth functions on $\Mcal$. If we embed a spacetime $\Mcal$ into another spacetime $\Ncal$, the field configurations on $\Ncal$ can be pulled back to 
$\Mcal$, and we obtain a functor $\mathfrak{E}$ from $\Loc$ to the category $\Vect$ of locally convex vector
spaces
\be
\mathfrak{E}(\Mcal)=\mathcal{C}^{\infty}(M,\RR)\ ,\ \mathfrak{E}\chi=\chi^*
\ee  
with the pullback $\chi^*\ph=\ph\circ\chi$ for $\ph\in\mathcal{C}^{\infty}(M,\RR)$. Note that $\mathfrak{E}$ is contravariant (i.e. it reverses the direction of the arrows), whereas the functor $\mathfrak{D}$ of test function spaces is covariant. The space $\mathfrak{E}(\Mcal)$ is obviously a vector space, but more generally it can also be a non-trivial infinite dimensional manifold modeled on $\mathfrak{D}(\mathcal{M})$ (see \cite{r11586,Michor,Neeb} for more details on infinite dimensional manifolds). Let us define the neighborhood basis $U+\ph$, where $\ph\in\E(\Mcal)$ and $U$ is an open neighborhood of $0$ in $\D(\Mcal)$, equipped with its standard inductive limit topology. The topology induced on $\E(\Mcal)$ by this basis is denoted by $\tau_W$. We can now define the atlas
\[
\{(\D(\Mcal)+\ph,\kappa_\ph)|\ph\in\E(\Mcal)\}\,,
\]
where coordinate charts $\kappa_\ph:\ph+\D(\Mcal)\rightarrow\D(\Mcal)$ are defined by $\kappa_\ph(\ph+\vec{\ph})=\vec{\ph}$. The coordinate change is given by $\kappa_{\ph_2}\circ\kappa_{\ph_1}^{-1}(\vec{\ph}_1)=\vec{\ph}_1+(\ph_1-\ph_2)$, where two charts overlap only if $\ph_1-\ph_2\in\D(\Mcal)$. This defines on $\E(\Mcal)$ a smooth affine manifold structure, in the sense of \cite{r11586}. Note, however, that $\E(\Mcal)$ equipped with $\tau_W$ is not a topological vector space (addition is not continuous). A similar construction can be applied also to situations in which the configuration space is not a vector space to start with, but still can be made into an infinite dimensional affine manifold (for example the space of all Lorentzian metrics, the space of gauge connections). The tangent space to  $\E(\Mcal)$ is  $T\E(\Mcal)=\E(\Mcal)\times \D(\Mcal)$ and the cotangent space is $T^*\E(\Mcal)=\E(\Mcal)\times \D'(\Mcal)$. Following \cite{BFR}, we can endow $T\E(\Mcal)$ with a flat connection.
Namely, vector fields are just smooth functions $X:\E(\Mcal)\to\D(\Mcal)$, 
and the covariant derivative of a vector field $X$ along another vector field $Y$ is defined by
\[
D_YX(\ph)\doteq\frac{d}{d\la}\Big|_{\la=0}X(\ph+\la Y(\ph))\,.
\]
The curvature of $D$ is 0, which will be important later on.

The classical observables are real valued functions on $\mathfrak{E}(\Mcal)$, i.e. (not necessarily linear) functionals. An important property of a functional is its spacetime support. It is defined as a generalization of the distributional support, namely as the set of points $x\in M$ such that $F$ depends on the field configuration in any neighbourhood of $x$.
\begin{align}\label{support}
\supp\, F\doteq\{ & x\in M|\forall \text{ neighborhoods }U\text{ of }x\ \exists \ph,\psi\in\E(\Mcal), \supp\,\psi\subset U\,,
\\ & \text{ such that }F(\ph+\psi)\not= F(\ph)\}\ .\nonumber
\end{align}
Here we will discuss only compactly supported functionals. If the configuration space  $\mathfrak{E}(\Mcal)$ is not a vector space but rather an infinite dimensional manifold (which will be the case in section \ref{gauge}), we replace the above definition with a bit more refined one, which uses the notion of the \textit{relative support}. Let $f_1,f_2$ be arbitrary functions between two topological spaces $X$ and $Y$, then
 \[
 \relsupp (f_1,f_2)\doteq \overline{\{x\in X| f_1(x)\neq f_2(x)\}}\,.
 \]
 Now we can provide a more general definition of the spacetime support of a functional on $\E(\Mcal)$:
 \begin{align}\label{support2}
\supp\, F\doteq\{ & x\in M|\forall \text{ neighbourhoods }U\text{ of }x\ \exists \ph_1,\ph_2\in\E(\Mcal), 
\\ & \relsupp(\ph_1,\ph_2)\subset U\,, \text{ such that }F(\ph_1)\not= F(\ph_2)\}\ .\nonumber
\end{align}

Next we want to select a class of functionals which are sufficiently regular such that all mathematical operations we want to perform are meaningful and which, on the other hand, is large enough to cover the interesting cases. One class one may consider is the class $\F_\reg(\Mcal)$ of regular polynomials defined as finite sums of functionals of the form
\be
F(\ph)=\int f_n(x_1,\dots,x_n)\ph(x_1)\dots\ph(x_n)d\mu_g(x_1)\ldots d\mu_g(x_n)\,,
\ee
where $f_n\in\mathcal{C}^{\infty}_c(M^n,\mathbb{R})$ and $d\mu_g$ is the volume form on $M$ induced by the metric $g$. Another important class, denoted by $\F_\loc(\Mcal)$, consists of the local functionals, i.e. functionals of the form
\be
F(\ph)=\int_M f(j^k(\ph))\,,
\ee
where $f$ is a density-valued function on the jet bundle and $j_x^k(\ph)\doteq(x,\ph(x),...,\ph^{\beta}(x))$ with  $\beta\in\NN_0^4$ , $|\beta|\leq k$ is the $k$-th jet of $\ph$ at $x\in M$. Note that the only regular polynomials in this class are the linear functionals, for example $F(\ph)=\int f(x)\ph(x)d\mu_g(x)$. Let $\F(\Mcal)$ denote the algebraic completion of   $\F_\loc(\Mcal)$ under the pointwise product
\be\label{pointwise}
(F\cdot G)(\ph)\doteq F(\ph)\cdot G(\ph)\,.
\ee
We call elements of $\F(\Mcal)$ \textit{multilocal functionals}. For later purposes (in particular for the quantization) we will need an even more general class of functionals than $\F(\Mcal)$. It turns out to be convenient to characterize this class of functionals in terms of their functional derivatives. The $n$th functional derivative $F^{(n)}$ of a functional $F$ is  
a compactly supported distributional density in $n$ variables, symmetrical under permutations of arguments, determined by
\[\langle F^{(n)}(\ph),\psi^{\otimes n}\rangle=\frac{d^n}{d\lambda^n}F(\ph+\lambda\psi)|_{\lambda=0}\,.\]
For a more precise definition and its relation to the smooth structure introduced by the topology $\tau_W$, see appendix \ref{smooth}.

We already have  all the kinematical structures we need. Now in order to specify a concrete physical model we need to introduce the dynamics. This can be done by means of 
a \textit{generalized Lagrangian}\index{generalized Lagrangian}. As the name suggests the idea is motivated by Lagrangian mechanics. Indeed, we can think of this formalism as a way to make precise the variational calculus
in field theory. Note that since our spacetimes are globally hyperbolic, they are never compact. Moreover we cannot restrict ourselves to compactly supported field configurations, since the nontrivial solutions of globally hyperbolic equations don't belong to this class. Therefore we cannot identify the action with a functional on $\E(\Mcal)$ obtained 
by integrating the Lagrangian density over the whole manifold. A standard solution to this problem is to integrate the  Lagrangian density only over a compact region of spacetime (for example over a time-slice, if $\Mcal$ has a compact Cauchy surface) and consider variations which vanish on the boundary of this compact region. This approach works for the classical theory, but in the process of quantization some spurious divergences appear. To avoid this problem, we follow \cite{BDF} and define a Lagrangian $L$ as a natural transformation between the functor of test function spaces $\D$ and the functor $\F_\loc$ such that it satisfies $\supp(L_\Mcal(f))\subseteq \supp(f)$ and the additivity rule 
\[
L_\Mcal(f+g+h)=L_\Mcal(f+g)-L_\Mcal(g)+L_\Mcal(g+h)\,,
\]
for $f,g,h\in\D(\Mcal)$ and $\supp\,f\cap\supp\,h=\varnothing$. We do not require linearity since in quantum field theory the renormalization flow does not preserve the linear structure; it respects, however, the additivity rule (see \cite{BDF}).

The action $S(L)$ is now defined as an equivalence class of Lagrangians  \cite{BDF}, where two Lagrangians $L_1,L_2$ are called equivalent $L_1\sim L_2$  if
\be\label{equ}
\supp (L_{1,\Mcal}-L_{2,\Mcal})(f)\subset\supp\, df\,, 
\ee
for all spacetimes $\Mcal$ and all $f\in\D(\Mcal)$. 
This equivalence relation allows us to identify Lagrangians differing by a total divergence.  For the free scalar field the generalized Lagrangian is given by
\be\label{Lscalar}
L_\Mcal(f)(\ph)=\frac{1}{2}\int\limits_M (\nabla_\mu\ph\nabla^\mu\ph-m^2\ph^2)fd\mu\,,
\ee
where $\nabla$ is the covariant derivative corresponding to $g$ and $m$ is the mass. The equations of motion are to be understood in the sense of \cite{BDF}. Concretely, the Euler-Lagrange derivative\index{derivative!Euler-Lagrange} of $S$ is a natural transformation $S':\E\to\D'$ defined as
\be\label{ELd}
\left<S'_\Mcal(\ph),h\right>=\left<L_\Mcal(f)^{(1)}(\ph),h\right>\,,
 \ee
for $f\equiv 1$ on $\supp h$. Note that for a fixed $\Mcal$ we have $S'_{\Mcal}\in\Gamma(T^*\E(\Mcal))$. The field equation is now the condition that
\be
 S_\Mcal'(\ph)=0\label{eom}\,,
\ee
so the space of solutions to this equation, denoted by $\E_S(\Mcal)$, is just the 0 locus of the 1-form $S'_{\Mcal}$. For the free scalar field we have (compare with section \eqref{freescalar}):
\[
 S_\Mcal'(\ph)=P\ph=-(\Box_g+m^2)\ph\,.
\]
 
We are now interested in characterizing the space $\F_S(\Mcal)$ of multilocal functionals on $\E_S(\Mcal)$. In the first step we can write it as the quotient
\[
\F_S(\Mcal)=\F(\Mcal)/\F_0(\Mcal)\,,
\]
where $\F_0(\Mcal)$ is the ideal in $\F(\Mcal)$ consisting of multilocal funcionals that vanish on $\E_S(\Mcal)$. 

In the next step we want to find a homological characterization of the quotient above. Note that if $X\in\Gamma(T\E(\Mcal))$ is a smooth vector field, then the functional $\left< S_\Mcal',X\right>$, obtained by contracting this vector field with the one-form $S_M'$, obviously vanishes on $\E_S(\Mcal)$. For this functional to be an element of $\F(\Mcal)$, we need some further regularity conditions on $X$. Let $\V(\Mcal)\subset\Gamma(T\E(\Mcal))$ denote the space of vector fields which are also derivations of $\F(\Mcal)$. We call such vector fields \textit{multilocal} and we observe that $\left< S_\Mcal',X\right>\in\F_0
(\Mcal)$, if $X$ is multilocal. Let us define a map $\delta_S:\V(\Mcal)\rightarrow\F(\Mcal)$ by 
\[\delta_S\doteq -\left< S_\Mcal',.\right>\,.
\]
The choice of sign is just a convention. Clearly, $\delta_S(\V(\Mcal))\subset\F_0(\Mcal)$. In general the opposite inclusion can hold only locally, since the structure of the global solution space of nonlinear PDE's can be veryt complicated. Since our ultimate goal is the quantum theory, we will avoid these complications by \textit{defining}, from now on, $\F_0(\Mcal)$ as $\delta_S(\V(\Mcal))$.

Under this identification, the space of functionals on the classical phase space is characterized as the 0-th homology of the following complex:
 \be\label{K}
\begin{array}{c@{\hspace{0,2cm}}c@{\hspace{0,2cm}}c@{\hspace{0,2cm}}c@{\hspace{0,2cm}}c@{\hspace{0,2cm}}c@{\hspace{0,2cm}}c@{\hspace{0,2cm}}c@{\hspace{0,2cm}}c}
\ldots&\rightarrow&\La_{\F}^2\V(\Mcal)&\xrightarrow{\delta_S}&\V(\Mcal)&\xrightarrow{\delta_S}&\F(\Mcal)&\rightarrow& 0\\
 &&2&&1&&0&&
\end{array}\ ,
\ee
where $\delta_S$ is extended to the algebra  $\Lambda_{\F}\V(\Mcal)$  of multilocal multi vector fields on $\mathfrak{E}(\mathcal{M})$ by requiring the graded Leibniz rule, acting from the right, with respect to the wedge product $\wedge$ of $\Lambda_{\mathfrak{F}}\mathfrak{V}(\mathcal{M})$. The space $\Lambda_{\F}\V(\Mcal)$ can be interpreted in terms of graded geometry. It is understood as  $\mathcal{O}_{\mathrm{ml}}(T^*[1]\mathfrak{E}(\mathcal{M}))$,  the algebra of multilocal functionals on $T^*[1]\E(\Mcal)$, the odd cotangent bundle of $\E(\Mcal)$. The term ``odd'' refers to the fact that the degree of the fiber is shifted by one with respect to the degree of the base (in this case the base is purely even, so the degree on the fiber, indicated in the square bracket,  is $+1$).  For a more general configuration space, that is $\mathfrak{E}(\mathcal{M})=\Gamma(E\rightarrow M)$ with some vector bundle $E$ over $M$, we have $T^*[1]\mathfrak{E}(\mathcal{M})=\mathfrak{E}(\mathcal{M})\oplus\mathfrak{E}^*(\mathcal{M})[1]$, where $\mathfrak{E}^*(\mathcal{M})\doteq \Gamma(E^*\rightarrow M)$ and $E^*$ is the dual bundle of $E$. Note that $\mathfrak{E}^*(\mathcal{M})$ can be embedded into $\Gamma_c'(E\rightarrow M)\equiv\mathfrak{E}_c'(\mathcal{M})$ using the paring defined as the integration with the volume form $d\mu_g$ on $M$. 

The precise meaning of $\mathcal{O}_{\textrm{ml}}$ is given in Ref. \cite{Book} and appendix A of Ref. 37. Broadly speaking, for a vector bundle $E_1$ over $M$, $\mathcal{O}_{\textrm{ml}}(\Gamma(E_1\rightarrow M)[1])$ is a subspace of 
\[
\prod_{n=0}^\infty \Gamma'((E_1)^{\boxtimes n}\rightarrow M^n)\,,
\]
consisting of sections satisfying appropriate WF set and symmetry properties. Here $\boxtimes$ denotes the exterior tensor product of vector bundles.

It is interesting to analyze the graded differential algebra $(\Lambda_{\F}\V(\Mcal),\delta_S)$ in more details and in particular to compute the higher cohomologies. Note that the kernel of $\delta_S$ in degree 1 consists of vector fields in $\V(\Mcal)$ which satisfy
\be\label{symm:vf}
\left<S'_\Mcal,X\right>\equiv 0\,.
\ee
We can interpret it as a condition that the directional derivative of the action $S_M$ along the vector field $X$ vanishes, i.e. the elements of $\ker\delta_S$ in degree 1 characterize the directions of constant $S_M$. This motivates the name \textit{local symmetries} for vector fields in $\V(\Mcal)$ which satisfy \eqref{symm:vf}. Among all the local symmetries there are in particular those which are in the image of $\delta_S$, i.e elements of  $\delta_S(\La_{\F}^2\V(\Mcal))$. Consider an example $X\wedge Y\in\Lambda_{\mathfrak{F}}^2\mathfrak{V}(\mathcal{M})$, where $X,Y\in\mathfrak{V}(\mathcal{M})$. In this case

\[\delta_S(X\wedge Y)=-(\delta_SX) Y+(\delta_S Y)X=\left<S_{\mathcal{M}}',X\right>Y-\left<S_{\mathcal{M}}',Y\right>X\,.\]
Such symmetries are called trivial, since they vanish on $\E_S(\Mcal)$ (as $S_{\mathcal{M}}'$ vanishes on $\mathfrak{E}_S(\mathcal{M})$). It follows that $H_1(\La_{\F}\V(\Mcal),\delta_S)$ is the space of local symmetries, modulo the trivial ones. If this homology vanishes, we say that the given theory possess no non-trivial local symmetries. This would be the case for the scalar field, but not for gauge theories. In fact, $H_1(\La_{\F}\V(\Mcal),\delta_S)$ is related to the hyperbolicity of linearised equations of motion \cite{FR}, the latter are induced by the second variational derivative of the action.

We come back to the discussion of the real scalar field. Note that the second variational derivative of the action induces a 2-form on $\D(\Mcal)$ . We define it in the following way:
\[
\left<S''_\Mcal(\ph),h_1\otimes h_2\right>\doteq \left<L^{(2)}_\Mcal(f)(\ph),h_1\otimes h_2\right>\,,
\]
where $f\equiv 1$ on the supports of $h_1$ and $h_2$. This defines $S''_\Mcal$ as a 2-form on $\E(\Mcal)$ (seen as a manifold equipped with the smooth structure induced by $\tau_W$) and $S''_\Mcal(\ph)$ induces an operator from $\D(\Mcal)$ to $\D'(\Mcal)$.
Actually, since $L_\Mcal(f)$ is local, the second derivative has support on the diagonal, so $S''_\Mcal(\ph)$ 
can be evaluated on smooth functions $h_1$, $h_2$, where only one of them is required to be compactly supported, so we obtain an operator (the so called linearized Euler-Lagrange operator) $P_{\varphi}:\E(\Mcal)\rightarrow \D'(\Mcal)$. 

We assume that, for all $\ph$, $P_{\varphi}$ maps in fact to $\E(\Mcal)\subset\D'(\Mcal)$ and that $P_{\varphi}:\E(\Mcal)\rightarrow \E(\Mcal)$ is a normally hyperbolic differential operator. We denote the retarded and advanced Green functions corresponding to this differential operator by $\Delta_S^R(\ph)$ and $\Delta_S^A(\ph)$, respectively. 
\subsubsection{Poisson structure}\label{Poisson}
In order to equip $\mathfrak{F}_S(\Mcal)$ with a Poisson bracket we rely on a method originally introduced by Peierls and consider the influence of adding a term to the action. Let $F\in\mathfrak{F}_\loc(\Mcal)$ be a local functional. We are interested in the flow $(\Phi_{\lambda})$ on $\mathfrak{E}(\Mcal)$ which deforms solutions of the original field equation $S_\Mcal'(\ph)=\beta$ with a given source term $\beta$ to those of the perturbed equation $S_\Mcal'(\ph)+\lambda F^{(1)}(\ph)=\beta$. Let $\Phi_0(\ph)=\ph$ and 
\be\label{flow}
\frac{d}{d\lambda}\left.\left(S_\Mcal'(\Phi_{\lambda}(\ph))+\la F^{(1)}(\Phi_{\lambda}(\ph))\right)\right|_{\lambda=0}=0 \ .
\ee
It follows that the vector field $\ph\mapsto X(\ph)=\frac{d}{d\lambda}\Phi_{\lambda}(\ph)|_{\lambda=0}$ satisfies the equation
\be\label{X:eq}
\left<S''_\Mcal(\ph),X(\ph)\otimes\cdot\right>+\left<F^{(1)}(\ph),\cdot\right>=0\,.
\ee
Equation \eqref{X:eq} can be written in a different notation as
\[
P_{\ph}X(\ph)+F^{(1)}(\ph)=0\,.
\]
We have two distinguished solutions for  $X$,
\be
X^{R,A}(\ph)=-\Delta_S^{R,A}(\ph)F^{(1)}(\ph)\ .
\ee
Let $\Delta_S(\ph)=\Delta^R_S(\ph)-\Delta^A_S(\ph)$ be the causal propagator. We introduce on $\mathfrak{F}_\loc(\Mcal)$ a product (the Peierls bracket)
\be\label{Peierls:bracket}
\Poi{F}{G}_S(\ph)=\langle F^{(1)}(\ph), \Delta_S(\ph)G^{(1)}(\ph)\rangle\,.
\ee

The Peierls bracket satisfies the conditions of a Poisson bracket, in particular the Jacobi identity. Note that if one of the entries of \eqref{Peierls:bracket} is in the ideal $\mathfrak{F}_0(\Mcal)$, also the bracket is in the ideal, hence the Peierls bracket induces a Poisson bracket on the quotient algebra $\F_S(\Mcal)$. Unfortunately $\F(\Mcal)$ is not closed under the Poisson bracket, so in order to obtain a Poisson algebra, one needs more singular functionals than the multilocal ones. Such a class of functionals arises naturally in the quantization, so we will postpone the problem of extending $\F(\Mcal)$ to a space closed under $\Poi{.}{.}_S$ to section \ref{backtoclass}.
\subsubsection{Example: Symplectic space of observables for the scalar field}\label{symplectic}
In this section we compare the abstract construction of the algebra of observables for the scalar field with other more familiar approaches, including the construction presented in section \ref{freescalar}. Let $\Mcal$ be a globally hyperbolic spacetime with a compact Cauchy surface. We restrict our attention to the vector space $\F_{\textrm{lin}}(\Mcal)$ consisting of functionals of the form
\be\label{Flin}
F_f(\ph)\doteq \int_M \ph(x)f(x)d\mu_g(x)\,,
\ee
where $f\in\D(\Mcal)$ and $d\mu_g$ is the invariant volume form on $\Mcal=(M,g)$. Among such functionals we can distinguish those which vanish on the solution space $\E_S(\Mcal)$, i.e. belong to the intersection of $\F_{\textrm{lin}}(\Mcal)$ with $\F_0(\Mcal)\cong\delta_S(\V(\Mcal))$. It is easy to see that they have to be of the form \eqref{Flin} with $f=Ph$ for $h\in\D(\Mcal)$, where $P$ is the Klein-Gordon operator. Let us denote the space of such functionals by $\F^0_{\textrm{lin}}(\Mcal)$ and the quotient $\F_{\textrm{lin}}(\Mcal)/\F^0_{\textrm{lin}}(\Mcal)$ by $\F^S_{\textrm{lin}}(\Mcal)$. The Peierls bracket induces on $\F^S_{\textrm{lin}}(\Mcal)$ a symplectic structure. By the argument above, $\F^S_{\textrm{lin}}(\Mcal)$ is isomorphic to the space of test functions modulo the image of the Klein-Gordon operator 
\[
\F^S_{\textrm{lin}}(\Mcal) \cong \D(\Mcal) / P \D(\Mcal)\,
\]
and it follows that the symplectic vector spaces $(\F^S_{\textrm{lin}}(\Mcal),\{.,.\}_S)$ and $(\Rcal,\sigma)$ introduced in \ref{freescalar} are isomorphic, since
\[
\{F_f,F_{h}\}_S=\int f(x)\Delta_S(x,y)h(y)d\mu_g(x)d\mu_g(y)=\sigma(f,h) \,.
\]
\subsubsection{Example: Canonical Poisson bracket of the interacting theory}
Let us now compare the Peierls bracket with the canonical Poisson bracket. In standard cases these two coincide. We illustrate this fact by considering the Lagrangian
\[
L_\Mcal(f)(\ph)=\int\limits_M \left(\frac{1}{2}\nabla_\mu\ph\nabla^\mu\ph-\frac{m^2}{2}\ph^2-\frac{\la}{4!}\ph^4\right)fd\mu_g\,.
\]
Then $S'_\Mcal(\ph)=-\left((\square+m^2)\ph+\frac{\lambda}{3!}\ph^3\right)$ and $S_\Mcal''(\ph)$ is the linear operator
\be
-\left(\square +m^2+\frac{\lambda}{2}\ph^2\right)
\ee
(the last term is to be understood as a multiplication operator). The Peierls bracket is
\be
\Poi{\Phi(x)}{\Phi(y)}_S=\Delta_S(\Phi)(x,y)\,,
\ee
where $\Phi(x)$, $\Phi(y)$ are evaluation functionals on $\E(\Mcal)$ (i.e $\Phi(x)(\ph)\doteq \ph(x)$)
and   $x\mapsto \Delta_S(\ph)(x,y)$ is a solution (at $\ph$) of the  linearized equation of motion with the initial conditions
\be
\Delta_S(\ph)(y^0,\mathbf{x};y)=0\ ,\ \frac{\pa}{\pa x^0}\Delta_S(\ph)(y^0,\mathbf{x};y)=\delta(\mathbf{x},\mathbf{y})\ .
\ee
This coincides with the Poisson bracket in the canonical formalism. Namely, let $\ph$ be a solution of the field equation.
Then
\be
0=\Big\{(\square +m^2)\Phi(x)+\frac{\lambda}{3!}\Phi^3(x),\Phi(y)\Big\}_
{can}(\ph)=(\square +m^2+\frac{\lambda}{2}\Phi(x)^2)\Big\{\Phi(x),\Phi(y)\Big\}_{can}(\ph)
\ee
hence the canonical Poisson bracket satisfies the linearized field equation with the same initial conditions as the Peierls bracket.  Therefore the two brackets coincide.
\subsection{Deformation quantization of free field theories}\label{deformation}
Let us assume that $\Delta_S$ doesn't depend on $\ph$, i.e. $S$ is at most quadratic. Starting from the algebra $\mathfrak{F}(\Mcal)$ equipped with the Poisson bracket $\Poi{.}{.}_S$ we set out to construct an associative product $\star$ on $\mathfrak{F}(\Mcal)[[\hbar]]$ such that for $\hbar\to0$
\be
F\star G\to F\cdot G 
\ee 
and
\be
[F,G]_{\star}/i\hbar\to \Poi{F}{G}_S \ .
\ee
Such construction is called formal deformation quantization. For the Poisson algebra of functions on a finite dimensional Poisson manifold, the formal deformation quantization exists  due to a theorem of Kontsevich \cite{Kon}. 
In field theory the formulas of Kontsevich lead to ill defined terms, and a general solution of the problem is not known. Nevertheless, in case the action is quadratic in the fields, the $\star$-product can be explicitly defined by
\be
(F\star G)(\ph)\doteq\sum\limits_{n=0}^\infty \frac{\hbar^n}{n!}\left<F^{(n)}(\ph),\left(\tfrac{i}{2}\Delta_S\right)^{\otimes n}G^{(n)}(\ph)\right>\,,
\ee 
which can be formally written as $e^{\frac{i\hbar}{2}\left\langle\Delta_S,\frac{\delta^2}{\delta\ph\delta\ph'}\right\rangle}F(\ph)G(\ph')|_{\ph'=\ph}$.
This product is well defined (in the sense of formal power series in $\hbar$) for regular functionals $F,G\in\F_\reg(\Mcal)$ and satisfies the conditions above. However, in order to describe more physically relevant situations, we want to extend the product to more singular functionals which include in particular the local ones. To this end we go over to an equivalent product by replacing $\frac{i}{2}\Delta_S$ by a distribution $W$ with the same antisymmetric part. This modification turns out to be related to the physical requirement of the positivity of energy. 

We want to find a split
\be
\tfrac{i}{2}\Delta_S=W-H
\ee
in such a way that $W$ should locally select positive frequencies, thus fulfilling the requirements on positivity of energy in the small. To give a mathematical meaning to this statement we need to use the concept of a wave front set.
{\begin{df}\label{WFset}
Let $t$ be a distribution on $\mathbb R^n$. The point $(x,k)\in\mathbb R^n\times(\mathbb R^n\setminus\{0\})$ is called regular if there exists a test function $f$ with $f(x)\neq0$ and a conical neighborhood $N$ of $k$ (i.e. $N$ is invariant under multiplication with a positive number) such that the Fourier transform of $tf$ decays faster than any polynomial within $N$. 

The {wave front set} of $t$, denoted $\mathrm{WF}(t)$, is defined as the complement (within $\mathbb R^n\times(\mathbb R^n\setminus\{0\})$) of the set of regular points.
\end{df}}
We want to use these facts for distributions on differentiable manifolds. The concept of a Fourier transform clearly depends on the choice of a chart, but fortunately the property that  the Fourier transform decays fast  in some direction has a geometrical meaning in terms of the cotangent space. Therefore,
the wave front set can be defined as a subset of the cotangent bundle $T^*M$. The form of the wave front sets of distributions occuring in quantum field theory is largely determined by the theorem on the propagation of singularities:

{\begin{thm}
Let $D$ be a differential operator with real principal symbol $\sigma_P(D)$ (here we follow the physicist's convention and identify $i\partial$ with a covector (momentum)).
Let $t$ be a distribution with $Dt=f$ with a smooth function $f$. Then the wave front set of $t$ is contained in the zero set of $\sigma_P(D)$ and is invariant under the Hamiltonian flow generated by $\sigma_P(D)$.
\end{thm}}
\begin{itemize}
\item For an elliptic operator the principal symbol vanishes only at the origin, hence solutions of elliptic equations have empty wave front sets and thus are smooth.

{\item For a normally hyperbolic operator the principal symbol coincides with the inverse metric. Therefore the covectors in the wave front sets of solutions are always lightlike, and the wave front set is invariant under the geodesic flux, i.e. with $(x,k)$ all points $(\gamma(t),g(\dot\gamma(t),\cdot))$ with $\gamma$ being the geodesic with initial condition $\gamma(0)=x$ and $g(\dot\gamma(0),\cdot)=k$ are elements of the wave front set.}
\end{itemize}
In particular, the WF set of $\Delta_S$ has the following form
\begin{equation}\label{spectrum0}
\WF(\Delta_S)=\{(x,k;x',-k')\in \dot{T}M^2|(x,k)\sim(x',k')\}\,,
\end{equation}
where $(x,k)\sim(x',k')$ means that there exists a null geodesic connecting $x$ and $x'$ and $k'$ is the parallel transport of $k$ along this geodesic. Selecting ``positive frequencies'' means that we require that the WF set of $W$ consists only of  covectors that are in the closed future lightcone, i.e.
\begin{equation}\label{spectrum}
\WF(W)=\{(x,k;x',-k')\in \dot{T}M^2|(x,k)\sim(x',k'), k\in (\overline{V}_+)_x\}\,.
\end{equation}
This motivates the following definition (after \cite{Rad}):
\begin{df}\label{Hadamard}
A bisolution $W$ of the Klein-Gordon equation is called a Hadamard function if
\begin{enumerate}

{\item \label{im} $2\mathrm{Im}\,W=\Delta_S$}

{\item it fulfills \eqref{spectrum} (microlocal spectrum condition ($\mu$SC))}

{\item \label{pos} $W$ is of positive type (hence $H$ is real).}
\end{enumerate}
\end{df}
Before the work of Radzikowski \cite{Rad}, Hadamard functions were characterized explicitly by their singularity structure. The characterization we will provide in definition \ref{locHad} follows the one presented in \cite{Hack} which is a local version of the global condition given by \cite{KW91}. {Radzikowski proved that the microlocal spectrum condition  \eqref{spectrum} is in fact equivalent to that global condition. 

We will need the explicit characterization of the singularity structure of Hadamard functions in the definition of covariant Wick powers which will follow at the end of this section, so we will now describe it in full detail. To this end we introduce some notation. Let $t:\Mcal\rightarrow\RR$ be a time function (smooth function with a timelike and future directed gradient field) and let
\[
\sigma_\varepsilon(x,y)\doteq \sigma(x,y)+2i\varepsilon(t(x)-t(y))+\varepsilon^2\,,
\]
where $\sigma(x,y)$ is half of the square geodesic distance between $x$ and $y$, i.e. 
\[
\sigma(x,y)\doteq \frac{1}{2}g(\exp_x^{\minus}(y),\exp_x^{\minus}(y))\,.
\]
\begin{df}\label{locHad}
We say that a bi-distribution $W$ on $\Mcal$ is of local Hadamard form if, for every $x_0\in M$, there exists a geodesically convex neighbourhood $V$ of $x_0$ such that, for every integer $N$, $W(x,y)$ on $V\times V$ can be written in the form
\be\label{Hadform}
W=\lim_{\varepsilon\downarrow0}\left(\frac{u}{\sigma_\varepsilon}+\sum_{n=0}^N\sigma^nv_n\log\left(\frac{\sigma_\varepsilon}{\la^2}\right)+w_N\right)=W_{N}^{\textrm{sing}}+w_N\,,
\ee
where $u, v_n \in\Ci(M^2,\RR)$, $n=0,\ldots,N$ are solutions of the transport equations and are uniquely determined by the local geometry, $\la$ is a free parameter with the dimension of inverse length and $w_N$ is an $2N + 1$ times continuously
differentiable function.
\end{df}

 On Minkowski space one could choose $W$ as the  Wightman 2-point-function, i.e. the vacuum expectation value of the product of two fields. This, however, becomes meaningless in a more general context, since a generally covariant concept of a vacuum state does not exist. Nevertheless, Hadamard functions exist on general globally hyperbolic spacetimes, as shown by \cite{FNW} (by means of the deformation argument).
 
It turns out that using $W$ instead of $\Delta_S$ in the definition of the star product allows to extend it to more singular functionals. To this end one applies H{\"o}rmander's criterion for multiplying distributions \cite{Hoer1}. Let $t$ and $s$ be distributions on $\mathbb R^n$. Let $f$ be a test function with sufficiently small support. We may multiply $f$ by another test function $\chi$ with slightly larger support which is identical to 1 on the support of $f$. Let $f_p(x)=f(x)e^{ipx}$ and $\chi_{-p}(y)=\chi(y)e^{-ipy}$.
The point-wise product of two distributions $t,s$ can be defined via the integral
\[\langle ts,f\rangle=\int \frac{d^np}{(2\pi)^n}\langle t,f_p\rangle\langle s,\chi_{-p}\rangle \] 
if the sum of the wave front sets,
{\[\{(x,k+k')|(x,k)\in\mathrm{WF}t,(x,k')\in\mathrm{WF}s\}\]}
{does not intersect the zero section $\{x,0)|x\in\mathbb R^n\}$.} This condition, called the  H{\"o}rmander's criterion, generalizes also to manifolds, since the construction uses only the local data.

Now we want to apply this criterion to our problem of extending the star product. For a given Hadamard function $W=\frac{i}{2}\Delta_S+H$ we consider the following linear functional differential operator on $\E(\Mcal)$:
\be
\Gamma_H=\left\langle H,\frac{\delta^2}{\delta\ph^2}\right\rangle
\ee
and define a new $\star$-product by
\be\label{Hprod}
F\star_HG=\al_H\left((\al^{-1}_HF)\star(\al_H^{-1}G)\right)\,, 
 \ee
 where 
 \[
 \al_H\doteq e^{\frac{\hbar}{2}\Gamma_H}\,.
 \]
 This product differs from the original one in the replacement of $\frac{i}{2}\Delta_S$ by $W$ and can be defined on a much larger space of functionals. H{\"o}rmander's criterion implies that a sufficient condition for $\star_H$ to be well defined is that we consider functionals which satisfy
 \be\label{mlsc}
\WF(F^{(n)}(\ph))\subset \Xi_n,\quad\forall n\in\NN,\ \forall\ph\in\E(\Mcal)\,,
\ee
where $\Xi_n$ is an open cone defined as 
\be\label{cone}
\Xi_n\doteq T^*M^n\setminus\{(x_1,\dots,x_n;k_1,\dots,k_n)| (k_1,\dots,k_n)\in (\overline{V}_+^n \cup \overline{V}_-^n)_{(x_1,\dots,x_n)}\}\,,
\ee
where $(\overline{V}_{\pm})_x$ is the closed future/past lightcone understood as a conic subset of
$T^*_xM$. We denote the space of smooth compactly supported functionals, satisfying (\ref{mlsc}) by $\F_\mc(\Mcal)$ and call them \textit{microcausal functionals}. This includes in particular the local functionals. For them the support of the functional derivatives is on the thin diagonal, and the wave front sets satisfy $\sum_i k_i=0$. 
 
The transition between two star products $\star$ and $\star_H$ corresponds to normal ordering, and the relation between them is an algebraic version of Wick's theorem. 
\begin{exa}
We list some important examples of $W$:
\begin{itemize}
\item On Minkowski space, we can choose $W$ as the positive frequency part $\Delta_+$ of $\Delta$ (the Wightman 2-point function), given by
\[\Delta_+(x,y)=\frac{1}{(2\pi)^{3}}\int d^3\mathbf p\frac{e^{-i\omega(\mathbf p)(x^0-y^0)+i\mathbf p(\mathbf x-\mathbf y)}}{2\omega(\mathbf p)}\ ,\]
with $\omega(\mathbf p)=\sqrt{|\mathbf p|^2+m^2}$.
We can check explicitly that the square of $\Delta_+$ is well defined.
\[
\int d^4x \Delta_+(x)^2f(x)=\frac{1}{(2\pi)^{3}}\int \frac{d^3\mathbf p}{2\omega(\mathbf p)}\int \frac{d^3{\mathbf q}}{2\omega(\mathbf q)}\int d^4x\, e^{-i(\omega(\mathbf p)+\omega(\mathbf q)x^0+i(\mathbf p+\mathbf q)\mathbf x}f(x)=\]
 \[\int \frac{d^3\mathbf p}{2\omega(\mathbf p)}\int \frac{d^3{\mathbf q}}{2\omega(\mathbf q)}\tilde f(\omega(\mathbf p)+\omega(\mathbf q),\mathbf p+\mathbf q)\ .\]
This integral converges, since $\tilde f$ is fast decreasing.
\item Another example of $W$ is the 2-point function of KMS states (states satisfying the Kubo-Martin-Schwinger (KMS) condition for
thermal equilibrium) with positive temperature $\beta^{-1}$
\[
\Delta_\beta(x)=\frac{1}{(2\pi)^{3}}\int \frac{d^3\mathbf p}{2\omega(\mathbf p)}\left(\frac{e^{-i\omega(\mathbf p)x^0}}{1-e^{-\beta\omega(\mathbf p)}}+\frac{e^{i\omega(\mathbf p)x^0}}{e^{\beta\omega(\mathbf p)}-1}\right)e^{i\mathbf p\mathbf x}
\]
The KMS-condition follows from the identity $\Delta_\beta(x^0-i\beta,\mathbf x)=\Delta_\beta(-x^0,\mathbf x)$
\item On a generic static spacetime with $K\ge c>0$, we may choose
\[(Wf)_t=\int ds \frac{e^{-i\sqrt{K}(t-s)}}{2\sqrt{K}}f_s\]
This is the ground state with respect to the time translation symmetry of the spacetime. 
\item The ground state in Schwarzschild spacetime becomes singular at the horizon, but the KMS state with Hawking temperature $\beta^{-1}=\frac{1}{4m}$ is regular at the horizon.
\end{itemize}
 \end{exa}
 
The map $\al_H$ provides the equivalence between $\star$ and $\star_H$ on the space of regular functionals $\F_\reg(\Mcal)$. Its image can be then completed to a larger space $\F_\mc(\Mcal)$. We can also build a corresponding (sequential) completion $\al_H^{-1}(\F_\mc(\Mcal))$ of the source space. This amounts to extending  $\F_\reg(\Mcal)$ with all elements of the form $\lim_{n\rightarrow \infty}\al_H^{-1}(F_n)$, where $(F_n)$ is a convergent sequence in $\F_\mc(\Mcal)$ with respect to the H\"ormander topology $\tau_\Xi$ (see Appandix \ref{Ht} and references \cite{BDF,BaerF,Hoer1} for details). To see how this abstract construction works in practice, let us consider the example of the Wick square (after \cite{FR12}):
 \begin{exa}\label{Wick}
Consider a sequence $F_n(\ph)=\int \ph(x)\ph(y)g_n(y-x)f(x)$ with a smooth density $f$ and a sequence of smooth densities $g_n$ which converges to $\delta(x-y) d\mu_g(x)$ in the H\"ormander topology ($\delta$ here denotes the Dirac distribution). By applying $\al_H^{-1}=e^{-\tfrac{\hbar}{2}\Gamma_H}$, we obtain a sequence 
\[
\al_H^{-1}F_n= \int (\ph(x)\ph(y)g_n(y-x)f(x)-\hbar H(x,y)g_n(y-x)f(x))\,,
\]
The limit of this sequence can be identified with $\int \mathopen:\ph(x)^2\mathclose:_Hf(x)$, i.e.:
\[
\int \mathopen:\ph(x)^2\mathclose:_Hf(x)=\lim_{n\rightarrow\infty}\int (\ph(x)\ph(y)-\hbar H(x,y))g_n(y-x)f(x)
\]
We can write it in a short-hand notation as a coinciding point limit:
\[
\mathopen:\ph(x)^2\mathclose:_H\,=\lim_{y\rightarrow x}(\ph(x)\ph(y)-\hbar H(x,y))\,.
\]
 We can see that transforming with $\al_H^{-1}$ corresponds formally to a subtraction of $\hbar H(x,y)$.
Now, to recognize Wick's theorem let us consider a product of two Wick squares $ \mathopen:\ph(x)^2\mathclose:_H \mathopen:\ph(y)^2\mathclose:_H$  and we drop the subscript $H$ to simplify the notation. With the use of the isomorphism $\al_H^{-1}$ this can be written as:
\begin{multline*}
\int \ph(x)^2 f_1(x)\star_H\int \ph(y)^2 f_2(y)=\\
\int \ph(x)^2 \ph(y)^2f_1(x) f_2(y)+2i\hbar \int \ph(x) \ph(y)W(x,y)f_1(x) f_2(y)-\hbar^2\int (W(x,y))^2f_1(x) f_2(y)\,.
\end{multline*}
Omitting the test functions and using $\al_H^{-1}$ we obtain
\[
\mathopen:\ph(x)^2\mathclose:\mathopen:\ph(y)^2\mathclose:=\mathopen:\ph(x)^2 \ph(y)^2\mathclose:+4 \mathopen:\ph(x) \ph(y)\mathclose:\frac{i\hbar}{2}W(x,y)+2\Big(\frac{i\hbar}{2}W(x,y)\Big)^2\,,
\]
which is a familiar form of Wick's theorem applied to $\mathopen:\ph(x)^2\mathclose:\mathopen:\ph(y)^2\mathclose:$.
\end{exa}
Analogous to \eqref{Hprod}, for two choices of a star product, $\star_H$ and $\star_{H'}$, we have a map $\al_{H-H'} $ that relates them. Since $H-H'$ is smooth, this map is an isomorphism, so all the algebras $(\al_H^{-1}(\F_\mc(\Mcal)[[\hbar]]),\star)$ are isomorphic and define an abstract algebra $\fA(\Mcal)$. Since for $F \in \fA(\Mcal)$ we have $\al_HF\in \fA^{\sst H}(\Mcal)\doteq\F_{\mc}(\Mcal)[[\hbar]],\star_H)$, we can realize $\fA(\Mcal)$ more concretely as the space of families $\{ \al_HF \}_H$,  labeled by possible choices of $H$, where  $F \in \fA(\Mcal)$, fulfilling the relation
\[
 F_{H'} = \exp(\hbar \Gamma_{H'-H}) F_H\,,
\]
equipped with the product
\[
 (F \star G)_H = F_H \star_H G_H.
\]
The support of $F \in \fA(\Mcal)$ is defined as $\supp(F) = \supp(\al_HF)$. Again, this is independent of $H$. Functional derivatives are defined by
\[
\left<F^{(1)}(\ph),\psi\right> = \al_H^{-1}\left<(\al_HF)^{(1)}(\ph),\psi\right>\,,
\]
which makes sense, since $\Gamma_{H'-H}$ commutes with functional derivatives. Note that the complex conjugation satisfies the relation:
\be
\overline{F\star G}=\overline{G}\star\overline{F}\,.
\ee
Therefore we can use it to define an involution  $F^*(\ph)\doteq\overline{F(\ph)}$. The resulting structure is an involutive noncommutative algebra $\fA(\Mcal)$, which provides a quantization of $(\F_{\mc}(\Mcal),\Poi{.}{.}_S)$. Any linear functional $\omega_{H,\ph}$ given by
 \be\label{vacuum}
 \om_{H,\ph}(F)=(\al_H(F))(\ph)\,,
 \ee 
 where $\ph\in\E_S(\Mcal)$, is a state on $\fA(\Mcal)$. The kernel of the associated GNS representation  (the representation provided by the Gelfand-Naimark-Segal
 (GNS) construction) is the ideal generated by the field equation.

Let us now compare the quantization procedure that we outlined above with the Weyl quantization on the example of the free scalar field performed in \ref{freescalar}. Consider the subalgebra of $\fA(\Mcal)$ generated by functionals of the form $\Wcal(f)\doteq\exp(iF_f)$, where $F_f\in \F_{\textrm{lin}}(\Mcal)$ is of the form \eqref{Flin}. For such generators we have:
\begin{align*}
\left< (\Wcal(f))^{(1)}(\ph),h\right>&= \frac{d}{d\lambda}\left( \Wcal(f) (\ph + \lambda h)\right)|_{\lambda=0}= \frac{d}{d \lambda}  e^{i \int  f (\ph + \lambda h)d\mu_g} \big|_{\lambda=0}=\\
&=\left( i \int f h\,d\mu_g\right) \Wcal(f)(\ph).\\
\end{align*}
\[
\left<(\Wcal(f))^{(n)}(\ph),h^{\otimes n}\right>=  \left( i \int f h\,d\mu_g\right)^n \Wcal(f)(\ph).
\]
Inserting this into the $\star$-product formula, we find,
\begin{eqnarray*}
\Wcal(f) \star \Wcal(\tilde{f}) &=&  \sum_{n=0}^{\infty} \left( \frac{i \hbar}{2}\right)^n \frac{(-1)^n}{n!} \left( \int \Delta_S(x,y) \tilde{f}(y)f(x)d\mu_g(x)d\mu_g(y)\right)^n\Wcal(f+\tilde{f})\\
&=& e^{-\frac{i \hbar}{2}\sigma(f,\tilde{f})} \Wcal(f+\tilde{f}),
\end{eqnarray*}
which are precisely the Weyl relations \eqref{Weyl}. One obtains the Weyl algebra after taking the quotient by the ideal generated by elements of the form $\exp(iF_{Ph})-1$, where $h\in\D(\Mcal)$.

 Let us now discuss the covariance properties of Wick products. As seen in example \ref{Wick}, polynomial functionals in $\fA^{\sst H}(\Mcal)$ can be interpreted as Wick polynomials.
Corresponding elements of $\fA(\Mcal)$ can be obtained by applying $\al_H^{-1}$. The resulting object will be denoted by
\be\label{polynomials1}
\int :\Phi_{x_1}\dots\Phi_{x_n}:_H f(x_1,\dots, x_n)\doteq \al^{-1}_H\Big(\int \Phi_{x_1}\dots\Phi_{x_n} f(x_1,\dots, x_n)\Big)\,.
\ee
where $\Phi_{x_i}$ are evaluation functionals and $f$ is an element of $\Ecal'_{\Xi_n}(M^n,\RR)$, the space of distributions on $M^n$ with the WF set contained in the open cone $\Xi_n$ defined in \eqref{cone}.
 
Now we want to construct Wick powers in such a way that they become elements of  $\Fcal_q$, the space of locally covariant quantum fields defined in section \ref{axioms}. On each object $\Mcal$ we have to construct the map ${\TT}_\Mcal$ from the classical algebra $\F_\loc(\Mcal)$ to the quantum algebra $\fA(\Mcal)$ in such a way that
\be
\label{covariance}
{\TT}_{\Mcal}(\Phi_{\Mcal}(f))(\chi^*\ph)={\TT}_{\Ncal}(\Phi_{\Ncal}(\chi_*f))(\ph)\,.
\ee
As we noted above, classical functionals can be mapped  to $\fA(\Mcal)$ by identification \eqref{polynomials1}. This however doesn't have right covariance properties, because $H$ cannot be chosen across all the globally hyperbolic spacetimes in a coherent way. A detailed discussion is presented in section 5 of \cite{BFV}, where it was shown that redefining Wick powers to become covariant amounts to solving a certain cohomological problem. To get a more explicit description, let us first consider a formal expression 
\[
A(x)(\ph)=e^{\la p(\nabla)\ph(x)}\,,
\]
where $p(\nabla)$ is a polynomial in covariant derivatives. We obtain a large class of elements of $\fA^{\sst H}_{\loc}(\Mcal)$ by using the prescription
\[
\Phi^{A,n}_{\Mcal}(f)(\phi)\doteq\int f(x)\frac{d^n}{d\la^n}A(x)(\ph)\big|_{\la=0}d\mu_g(x)\,.
\]
We now set
\[
A_H(x)=e^{\frac12p(\nabla)\otimes p(\nabla)w_N(x,x)}A(x)\,,
\]
where $w_N$ is an $2N + 1$ times continuously
differentiable function from definition \ref{locHad} and $N$ has to be larger than or equal to twice the degree of $p$, and find
\[
\al_{H-H'}A_{H'}(x)=A_H(x)\,,
\]
so covariant expressions can be obtained as
\[
\int f\,\frac{d^n}{d\la^n}A_H \big|_{\la=0}d\mu_g\,.
\]
Hence locally covariant Wick powers are given by
\[
:\Phi^{A,n}_{\Mcal}:(f)\equiv {\TT}_{\Mcal}(\Phi^{A,n}_{\Mcal}(f)) \doteq \al_H^{\minus} \int f\,\frac{d^n}{d\la^n}A_H \big|_{\la=0}d\mu_g\,.
\]
The result reproduces the solution, proposed earlier in \cite{HW}, to use $\al^{-1}_{H-w}$, where $w$ is the smooth part of the Hadamard 2-point function in formula \eqref{Hadform}. This is understood as $\lim_{N\rightarrow \infty} \al^{\minus}_{H-w_N}F$ for $F\in\F_{\loc}(\Mcal)$ and this limit makes sense, because the series converges after finitely many steps. The map ${\TT}_\Mcal:\F_\loc(\Mcal)\rightarrow\fA(\Mcal)$ can be now set as $\al^{-1}_{H-w}$ (which is fixed up to the choice of scale $\la$).  In \cite{HW}  ${\TT}_\Mcal$ is first characterized giving the list of axioms. The definition provided above satisfies these axioms and any other solution differs only by modifying the Wick polynomials by adding locally covariant fields of lower order.
\subsection{Back to the classical theory}\label{backtoclass}
We are now ready to answer the question posed at the end of section \ref{Poisson}: what is a good extension of the space $\F(\Mcal)$ which is closed under the Peierls bracket $\Poi{.}{.}_S$? It turns out that the class of microcausal functionals $\F_\mc(\Mcal)$, which we have introduced for the purpose of defining the star product in quantum theory fits very well this purpose. The Peierls bracket is well defined on the whole of $\F_\mc(\Mcal)$, provided $\Delta_S$ is independent of $\ph$. If this is not the case, then a stronger condition is needed. We define $\overline{\F}(\Mcal)$ to be a subspace of $\F_{\mc}(\Mcal)$  consisting of functionals $F$, such that the first derivative $F^{(1)}(\ph)$ is a smooth section for all $\ph\in\E(\Mcal)$ and 
$\ph\mapsto F^{(1)}(\ph)$ is smooth as a map $\E(\Mcal)\rightarrow \E(\Mcal)$, where $\E(\Mcal)$ is equipped with the compact open topology. It was shown in \cite{BFRej13} (Appendix A) that the pair $\fP(\Mcal)\doteq(\overline{\F}(\Mcal)),\{.,.\}_S)$ indeed forms a Poisson algebra defining the classical theory.
One can also check that $\fP$ is a covariant functor from $\Loc$ to $\Po$, the category of Poisson algebras. By taking the quotient by the ideal of functionals that vanish on the solution space, one obtains a symplectic space $\fP_S(\Mcal)\doteq(\overline{\F}_S(\Mcal)),\{.,.\}_S)$.
\subsection{Interacting theories and the time ordered product}
In the next step of our construction of LCQFT models we want to introduce the interaction. This is done in the perturbative way, so for a given action $S$ we first need to split it into the quadratic part and the interaction term. We choose a point in the configuration space $\ph_0\in\E(\Mcal)$ and split
\be
L_\Mcal(f)(\ph_0+\psi)=\frac12 \left< L(f)^{(2)}_\Mcal(\ph_0),\psi\otimes\psi\right>+{L_I}_\Mcal(f)(\ph_0,\psi) \ .
\ee
The first term is quadratic in fields and it induces an action which we denote by $S_0$. From now on we will work in a fixed spacetime $\Mcal$, so we drop the subscript $\Mcal$ in $S_0$ and $S_I$, not to overburden the notation. We have $S=S_0+S_I$ and we assume that $S_0''$ induces a normally hyperbolic operator for which retarded and advanced Green's functions exist. To introduce an interaction, one would like to use the interaction picture of quantum mechanics. 

For the moment we restrict to Minkowski space and we choose $\ph_0=0$. We consider the standard Fock space representation of the free scalar field. Let $H_0$ be the Hamiltonian operator of the free theory and let  $H_I=-\int  \Lcal_I(0,\mathbf x)d^3\mathbf x$ be the interaction Hamiltonian, where $\Lcal_I$ is the normal-ordered Lagrangian density (related to the classical quantity ${L_I}$). The time evolution operator in the interaction picture can be obtained by the Dyson formula as a time ordered exponential

{\[U(t,s)=e^{itH_0}e^{-i(t-s)(H_0+H_I)}e^{-isH_0}=\]
\[1+\sum_{n=1}^\infty\frac{i^n}{n!}\int_ {([s,t]\times\mathbb R^3)^n}d^{4n}xT\Lcal_I(x_1)\dots \Lcal_I(x_n)\]}
{with the time ordered powers of the operator valued function
\[x\mapsto \Lcal_I(x)=e^{iH_0x^0}\Lcal_I(0,\mathbf x)e^{-iH_0x^0}\,.
\]
{A typical example for an interaction is $\Lcal_I(x)=:\ph(x)^4:$}

Unfortunately this way to introduce the interaction leads to severe problems. In 4 dimensions typical Lagrangian densities (for example $\ph^4$) cannot be restricted to the time zero hyperplane as operator valued distributions, hence the time ordered products are not well defined (UV problem). Moreover, the integration over the interval $[s,t]$ amounts to an application of a distribution to a discontinuous function and leads to additional divergences (the so-called St\"uckelberg divergences). Another problem is related to the fact that the integral over all $\mathbf x$ does not exist (problem of the adiabatic limit). Finally, the overall sum might not converge.

{The UV problem can be solved by perturbative renormalization. The St\"uckelberg divergences can be avoided by a clever reformulation in which we replace the characteristic function of the interval $[s,t]$ with a smooth function. The problem of the adiabatic limit can be solved on the level of the time evolution of observables. The final problem with the convergence of the sum is presently out of reach in 4 dimensions.}

We want to use the time evolution operators to define interacting fields.
{We try first the following ansatz: Let $s<x^0<t$. Then
\[\ph_I(x)=U(x^0,s)^{-1}\ph(x)U(x^0,s)=U(t,s)^{-1}U(t,x^0)\ph(x)U(x^0,s)\]
formally solves the problem.
Note that $\ph_I(x)$ does not depend on the choice of $t>x^0$, and that a change of $s$ to another $s'<x_0$ amount to a unitary tranformation by $U(s,s')$. However, due to the mentioned problems, the formula above is only heuristic. In order to make it meaningful, we perform the following steps :

{\begin{enumerate}
\item We replace the integral over $\mathbf x$ by an integration over a test density $g_1$. In view of the finite propagation speed in local relativistic theories this does not change the time evolution in a finite region if the test density equals $d^3\mathbf x$ in a sufficiently large region. 
\item Replace the instantaneous switching on and off of the interaction by a smooth time dependence $H_I(t')g_2(t')$ with a test density $g_2$ with compact support. If $g_2$ is equal to $dt$ on the intervall $[s,t]$ we get again a solution of the field equation.

We now smear both in space and time and set for a test density $g$ on $\RR^4$
\[S(g)=1+\sum_{n=1}^\infty\frac{i^n}{n!}\int g(x_1)\dots  g(x_n)T\Lcal_I(x_1)\dots \Lcal_I(x_n)\]
and call this object the formal S-matrix. $S(g)$ is well defined if the time ordered product of $\Lcal_I$ is known.
\item There remains the singularity at $x^0$. However, we only need to define the interacting field as a distribution. Hence we integrate with  a test density $f$ to obtain
\[\int f(x)\ph_I(x)=\]
\[S(g)^{-1}\sum_{n=0}^\infty \frac{i^n}{n!}\int f(x)g(x_1)\dots g(x_n)T\ph(x)\Lcal_I(x_1)\dots \Lcal_I(x_n)\]
\[=\frac{d}{d\lambda}|_{\lambda=0}S(g)^{-1}S(g,\lambda f)\] 
where $S(g,f)$ is the formal S-matrix with the interaction density $g\Lcal_I+f\ph$ (Bogoliubov's formula). 
\end{enumerate}

Note that $\int f(x)\ph_I(x)$ does not depend on the behavior of $g$ outside of the past of $\mathrm{supp}f$ and depends on the behavior outside of the future of $\mathrm{supp}f$ via a unitary transformation independent of $f$. The limit of $S(g)$ for $g\to d^4x$ yields Dyson's formula for the S-matrix (adiabatic limit).

These considerations are the starting point for the program of causal perturbation theory (\cite{SR,BS,BP,EG}).
{It essentially consists in defining the time ordered products of Wick products of free fields as operator valued distributions on Fock space.}
{The time ordered products are required to satisfy a few axioms, the most important one being that the time ordered product coincides with the operator product if the arguments are time ordered.}
{Epstein and Glaser \cite{EG} succeeded in proving that solutions satisfying the axioms exist and that the ambiguity is labeled by the known renormalization conditions. The solution can be either constructed directly or via one of the known methods (BPHZ  (Bogoliubov-Parasiuk-Hepp-Zimmermann), Pauli-Villars, momentum cutoff and counter terms, etc.).}

After these heuristic considerations we return to the general problem of constructing interacting quantum field theories on curved spacetime. Let us define the Dirac propagator (at $\ph_0$) by $\Delta_S^D\doteq\frac12(\Delta_S^R(\ph_0)+\Delta_S^A(\ph_0))$. We can now introduce the linear operator
\be
\TT=e^{i\hbar\langle\Delta_S^D,\frac{\delta^2}{\delta\psi^2}\rangle} 
\ee
which acts on $\F_\reg(M)$ as
\[
(\TT F)(\ph)\doteq \sum_{n=0}^\infty \frac{\hbar^n}{n!}\left<(i\Delta_S^D)^{\otimes n},F^{(2n)}(\ph)\right>\,.
\]
By
\be\label{Tprod}
F\T G=\TT\left(\TT^{-1}F\cdot \TT^{-1}G\right)
\ee
we define a new product on $\mathfrak{F}_{\reg}(M)$ which is called \textit{the time-ordered product}. It is associative, commutative and can be seen as the time-ordered version of the $\star$-product, since
\be\label{ordering}
F\T G=\left\{\begin{array}{rcl}
F\star G&\textrm{if}&\supp G\prec\supp F\,,\\
G\star F&\textrm{if}&\supp F\prec\supp G\,,
\end{array}\right.
\ee
where the relation ``$\prec$'' is to be understood as ``not later than'' i.e. there exists a Cauchy surface which separates $\supp G$ and $\supp F$ and in the first case  $\supp F$ is in the future of this surface and in the second case it's in the past.

Note also that $\T$ is equivalent to the pointwise product of the classical theory by means of \eqref{Tprod}. Using the time-ordered product we can define the formal $S$-matrix $\Scal:\F_\reg(\Mcal)\rightarrow\F_\reg(\Mcal)[[\hbar]]$ as the time ordered exponential:
\be\label{Smatrix}
\Scal(V)\doteq e_{\sst{\TT}}^{iV/\hbar}=\TT\big( e^{\TT^{-1}iV/\hbar}\big) \,.
\ee
Eventually we would like to be able to define $\Scal$ for local arguments, since $V$ plays here the role of the interaction functional and non-trivial local interactions cannot be regular. We will deal with this problem in the next section. For the moment we put this issue aside and work with regular functionals only. Interacting quantum fields are constructed from the free ones by means of the Bogoliubov formula \cite{BS}:
\[
R_V(F)\doteq -i\hbar\frac{d}{d\la}\Scal_{V}(\la F)\Big|_{\la=0},
\]
where $V$ is the interaction term and $\Scal_{V}$ is the relative S-matrix defined by
\[
\Scal_{V}(\la F)\doteq \Scal(V)^{\star\minus}\star\Scal(V+F)\,,
\]
i.e.
\be\label{RV}
R_V(F)=\left(e_{\sst{\TT}}^{i V/\hbar}\right)^{\star\minus}\star\left(e_{\sst{\TT}}^{i V/\hbar}\T F\right)\,,
\ee
We interpret $R_V(F)$ as the interacting quantity corresponding to $F$.
\global\long\def\poisson#1#2{\left\lfloor #1,#2\right\rceil }

\global\long\def\bld#1{\boldsymbol{#1}}

\begin{fmffile}{SettingPAQFT}

\def\FD{\parbox{7mm}{
\begin{center}
\begin{fmfgraph}(5,5)
\fmfleft{F}
\fmfdot{F}
\end{fmfgraph}
\end{center}}}

\def\FFlop{\parbox{20mm}{
\begin{center}
\begin{fmfgraph}(20,15)
\fmfleft{F} \fmfright{G}
\fmfdot{F}
\fmf{phantom}{F,G}
\fmf{plain}{F,F}
\end{fmfgraph}
\end{center}}}

\def\FG{\parbox{13mm}{
\begin{center}
\begin{fmfgraph}(20,15)
\fmfleft{F} \fmfright{G}
\fmfdot{F,G}
\fmf{phantom}{F,G}
\end{fmfgraph}
\end{center}}}

\def\dumbbell{\parbox{10mm}{
\begin{center}
\begin{fmfgraph}(20,10)
\fmfleft{F} \fmfright{G}
\fmfv{decor.shape=circle,decor.filled=empty, decor.size=2thick}{F,G}
\fmf{plain}{F,G}
\end{fmfgraph}
\end{center}}}

\def\FoneFG{\parbox{20mm}{
\begin{center}
\begin{fmfgraph}(20,15)
\fmfleft{F} \fmfright{G}
\fmfdot{F,G}
\fmf{phantom}{F,G}
\fmf{plain}{F,F}
\end{fmfgraph}
\end{center}}}

\def\FGoneG{\parbox{20mm}{
\begin{center}
\begin{fmfgraph}(20,15)
\fmfleft{F} \fmfright{G}
\fmfdot{F,G}
\fmf{phantom}{F,G}
\fmf{plain}{G,G}
\end{fmfgraph}
\end{center}}}

\def\FoneG{\parbox{13mm}{
\begin{center}
\begin{fmfgraph}(20,15)
\fmfleft{F} \fmfright{G}
\fmfdot{F,G}
\fmf{plain}{F,G}
\end{fmfgraph}
\end{center}}}

\def\FdecoG{\parbox{15mm}{
\begin{center}
\begin{fmfgraph*}(20,15)
\fmfleft{F} \fmfright{G}
\fmflabel{$F$}{F}
\fmflabel{$G$}{G}
\fmfdot{F,G}
\fmf{fermion}{F,G}
\end{fmfgraph*}
\end{center}}}
\subsection{Renormalization}
Up to now we gave the formulas for the formal S-matrix and the interacting fields only in the situation where all the functionals are regular. This is very restrictive and doesn't cover most of the physically interesting cases, since interaction terms and observable encountered in QFT are usually local and local functionals can be regular if they are at most linear in the fields. It is natural to ask the question if formulas \eqref{Smatrix} and \eqref{RV} extend to local, non-linear arguments. An easy extension would be provided by the operation of normal ordering as described in section \ref{deformation}. This operation transforms the time ordering operator $\TT$ into another one $\TT^{\sst H}$, such that the new time ordered product is now defined with respect to the Feynman propagator $\Delta_S^F=i\Delta_S^D+H$, no longer the Dirac propagator $\Delta_S^D$. The operator $\mathcal{T}^H$ is then formally interpreted as the convolution with the oscillating Gaussian measure $i\hbar\Delta_S^F$, i.e.
\[
\mathcal{T}F(\varphi)\stackrel{\mathrm{formal}}{=}\int F(\varphi-\phi)d\mu_{i\hbar\Delta_S^F}(\phi)\,.\] 
Note that the Feynman propagator does depend on the choice of $H$. On Minkowski spacetime there is a natural choice of $H=\Delta_1$, related to the existence of the uniques vacuum state with the 2-point function $W=\frac{i}{2}\Delta_S+\Delta_1$. Contrary to the $\star_H$ product which is everywhere defined due to the wave front set properties of the positive frequency part of $\Delta_S$, the time ordered product is in general undefined since the wave front set of the Feynman propagator contains the wave front set of the $\delta$-function. However, there is a way to extend $\TT^{\sst H}$ to \textit{local} functionals. Let 
\[
\TT^{\sst H}_n(F_1,\dots,F_n)\doteq F_1{\cdot_{\TT_H}}\dots {\cdot_{\TT_H}} F_n,
\]  
if it exists. Here $F_1,\dots,F_n$ are elements of $\mathfrak{A}^H_{\mathrm{loc}}(\mathcal{M})\doteq(\mathfrak{F}_{\mathrm{loc}}(\mathcal{M})[[\hbar]],\star_H)$ and  ${\cdot_{\TT_H}}$ is defined analogously to $\T$ in \eqref{Tprod}, but $\Delta_S^D$ is replaced with $\Delta_S^F$. 

We call $\TT^{\sst H}_n$ the $n$-th order time-ordered product. In particular, $\TT^{\sst H}_2$ is well defined for $F_1$, $F_2$ with disjoint supports, since the binary time-ordered product can be expressed in terms of the star product, according to \eqref{ordering}. This causality relation is the key property and can be iterated to obtain $\TT^{\sst H}_n(F_1,\dots,F_n)=\TT^{\sst H}_k(F_1,\dots,F_k)\star_H \TT^{\sst H}_{n-k}(F_{k+1},\dots,F_n)$ if the supports $\supp F_i$, $i=1,\dots,k$ of the first $k$ entries do not intersect the past of the supports $\supp F_j$, $j=k+1,\dots,n$ of the last $n-k$ entries. This property, called the \textit{causal factorisation property} can be taken as an axiom that also the extended time-ordered products have to satisfy. This leads to the idea of Epstein and Glaser (\cite{EG}) to construct the time ordered products  $(\TT^{\sst H}_n)_{n\in\NN_0}$ as $n$-linear symmetric maps from local functionals (without the condition of disjoint supports) to $\F_\mc(\Mcal)$ by induction, assuming the following axioms:
\begin{enumerate}[{\bf T 1.}]
\item $\TT^{\sst H}_0=1$,\label{Ts}
\item  $\TT^{\sst H}_1=\id$,
\item $\TT^{\sst H}_n(F_1,\dots,F_n)=\TT^{\sst H}_k(F_1,\dots,F_k)\star_H \TT^{\sst H}_{n-k}(F_{k+1},\dots,F_n)$ if the supports $\supp F_i$, $i=1,\dots,k$ of the first $k$ entries do not intersect the past of the supports $\supp F_j$, $j=k+1,\dots,n$ of the last $n-k$ entries.\label{Te}
\end{enumerate}
One can leave some freedom in the definition of $\mathcal{T}_1^H$, which can then be used to absorb the ambiguity in defining normal-ordering. On the technical side, at each step $n$ of the induction one constructs $\TT^{\sst H}_n$ knowing $\TT^{\sst H}_k$, $k<n$ by extending certain distributions which are well defined everywhere, away from the thin diagonal of $M^n$ (see \cite{BF0,HW} for details). It turns out that the freedom to add a $n$-linear  map $
Z_n:\F_\loc(\Mcal)^n\to\F_\loc(\Mcal)[[\hbar]]$ which describes possible finite renormalizations. The renormalized S-matrix  for the interaction $\lambda V$ is defined by
\[
\Scal(\lambda V)=\sum_{n=0}^\infty \tfrac{\lambda ^n}{n!}\TT^{\sst H}_n(V,\ldots,V)\,,
\]
where $\lambda$ is another formal parameter, interpreted as the coupling constant. Summing up the maps $Z_n$ we obtain $Z:\F_\loc(\Mcal)[[\hbar]]\rightarrow\F_\loc(\Mcal)[[\hbar]]$ with the following properties:
\begin{enumerate}[{\bf Z 1.}]
\item $Z(0)=0$,\label{Zs}
\item $Z^{(1)}(0)=\id$,
\item $Z=\id+\Ocal(\hbar)$,
\item $Z(F+G+H)=Z(F+G)+Z(G+H)-Z(G)$, if $\supp\,F\cap\supp\,G$,
\item $\frac{\delta Z}{\delta\ph}=0$.\label{Zf}
\end{enumerate}
We call the group of formal diffeomorphisms of $\F_\loc(\Mcal)[[\hbar]]$ that fulfill {\bf Z\ref{Zs}}-{\bf Z\ref{Zf}} the  St{\"u}ckelberg-Petermann renormalization group $\Rcal$. The relation between the formal S-matrices and elements of $\Rcal$ is clarified by the main theorem of renormalization \cite{PS82,DF04,BDF}. It states if $\Scal$ and $\hat{\Scal}$ are S-matrices built from time ordered products  satisfying the axioms {\bf T\ref{Ts}}-{\bf T\ref{Te}}, then there exists $Z\in\Rcal$ such that
\be
\hat{\Scal}=\Scal\circ Z 
\ee
where $Z\in\Rcal$ and conversely, if $\Scal$ is an S-matrix satisfying the axioms {\bf T\ref{Ts}}-{\bf T\ref{Te}} and $Z\in\Rcal$ then also $\hat{\Scal}$ fulfills the axioms.

In \cite{FR3} it was shown that the renormalized time ordered product can be extended to an associative, commutative binary product defined on the domain $\Dcal_{\TT^{\sst H}}(\Mcal)\doteq\TT^{\sst H}(\F(\Mcal))$, where $\TT^{\sst H}\doteq\oplus_n\TT^{\sst H}_n\circ m^{-1}$.
Here $m^{-1}:\F(\Mcal)\to S^\bullet\F^{(0)}_\loc(\Mcal)$ is the inverse of the multiplication, as defined in \cite{FR3,Rej11b}. $\Dcal_{\TT^{\sst H}}(\Mcal)$ contains in particular $\fA^{\sst H}_\loc(\Mcal)$ and is invariant under the renormalization group action. We can now define $(\TT F)\doteq \TT^{\sst H}(F_H)$.
 Renormalized time ordered products are defined by
\be
A\T B\doteq\TT(\TT^{\minus}A\cdot\TT^{\minus}B)\,.
\ee
\subsection{Locally covariant interacting fields}
In this section we construct a local net of $*$-algebras corresponding to the interacting theory on a fixed spacetime $\Mcal$. For clarity, we will drop the subscript $\Mcal$ in $L_{\Mcal}$ not to overburden the notation. Let $\Ocal$ be a relatively compact open subregion of some spacetime $\Mcal$. We already know that for $F\in\F_{\loc}(\Ocal)$, $\Scal_{\lambda L_{I}(g)}(F)$ depends only on the behavior of $g$ within $J_-(\Ocal)$, but the dependence on $g$ in that part of the past which is outside of $J_+(\Ocal$) is via a unitary transformation which is independent of $F$. More concretely, if $g'$ coincides with $g$ on a neighborhood of $J^{\diamond} (\Ocal):= J_+(\Ocal) \cap J_-(\Ocal)$, then there exists a unitary $U(g', g) \in \fA[[\hbar,\lambda]]$  (formal power series in $\hbar$ and the the coupling constant $\lambda$) such that
\[
\Scal_{\lambda L_{I}(g')}(F)= U(g', g)\Scal_{\lambda L_{I}(g)}(F)U(g', g)^{-1}\,,
\]
for all $F\in\F_{\loc}(\Ocal)$. It follows that the algebra generated by the elements of the form  $\Scal_{\lambda L_{I}(g)}(F)$  is, up to isomorphy, uniquely determined by the restriction of $g$ to the causal completion $J^{\diamond} (\Ocal)$. This defines an abstract algebra $\fA_{\lambda L_I[g]}(\Ocal)$, where  $[g]\equiv[g]_\Ocal$ denotes the class of all test functions which coincide with $g$ on a neighborhood of $J^{\diamond}(\Ocal)$. One then can insert instead of $g$ a smooth function $G$ without restrictions on the support. The algebra $\fA_{\lambda L_I[G]}(\Ocal)$, is generated by maps
\[
R_{\lambda L_{I}[G]}(F):[G]_{\Ocal}\rightarrow \fA(\Mcal),\quad g\mapsto R_{\lambda L_{I}(g)}(F)=i\hbar\frac{d}{d\mu}\Scal_{\lambda L_{I}(g)}(\mu F)\Big|_{\mu=0}\,.
\]
Now if $\Ocal_1\subset \Ocal_2$, we can then define a map $\fA_{\lambda L_I[g]}(\Ocal_1)$ to $\fA_{\lambda L_I[g]}(\Ocal_2)$ by taking the restriction of maps $R_{\lambda L_{I}[G]_{\Ocal_1}}(F)$ to $[G]_{\Ocal_2}$. For $G=1$ we denote $\fA_{\lambda L_I[1]}(\Ocal)\equiv \fA_{\lambda S_I}(\Ocal)$ and analogously $R_{\lambda L_{I}[1]}(F)\equiv R_{\lambda S_{I}}(F)$ for $F\in\F_\loc(\Ocal)$. We can now construct the inductive limit $\fA_{\lambda S_I}(\Mcal)$ of the net of local algebras  $(\fA_{\lambda S_I}(\Ocal))_{\Ocal\subset\Mcal}$. We call this the \textit{algebraic adiabatic limit}. $\fA_{\lambda S_I}$ is a functor from $\Loc$ to $\Obs$ and for a locally covariant classical field $\Phi$ (a natural transformation from $\D$ to $\F$) we obtain a locally covariant interacting quantum field by taking $R_{\lambda S_I}(\Phi)_{\Mcal}(f)\doteq R_{\lambda S_I}(\Phi_{\Mcal}(f))$, provided all the time ordered products $\TT^H_n$ are constructed in a covariant way. The existence of such covariant time-ordered products was proven in \cite{HW}.
\section{Gauge theories}\label{gauge}
\subsection{Classical theory}
In section \ref{genLagr} we have already shown that, under some regularity assumptions, the space of on-shell functionals $\F_S(\Mcal)$ can be characterized as the 0th homology of the differential complex  $(\Lambda_{\F}\V(\Mcal),\delta_S)$ (see \eqref{K}). We have also indicated that if the first homology is non-trivial, then the theory possesses local symmetries. In this chapter we discuss in detail how to quantize such theories using the BV framework (named after Batalin and Vilkoviski \cite{BV81}), in the version proposed in \cite{FR,FR3}. First, we note that $(\Lambda_{\F}\V(\Mcal)$ as the space of multivector fields, is equiped with a natural structure of minus the Schouten bracket $\{.,.\}$. This is an odd Poisson bracket defined by the following properties:
\begin{enumerate}
\item $\{X,F\}\doteq-\partial_X F$, for $F\in\F(\Mcal)$ and $X\in\V(\Mcal)$,
\item $\{X,Y\}\doteq -[X,Y]$, for $X,Y\in\V(\Mcal)$,
\item For higher order multivector fields we extend $\{.,.\}$ by imposing the graded Leibniz rule:
\be\label{leibniz}
\{X,Y\wedge Z\}=\{X,Y\}\wedge Z+(-1)^{nm}\{X,Z\}\wedge Y\,,
\ee
where $n$ is the degree of $Y$ and $m$ the degree of $Z$. 
\end{enumerate}
Note that $\delta_S$ is locally generated by the bracket in the sense that
\[
\delta_S X=\{X,S_\Mcal\}\doteq\{X,L_{\Mcal}(f)\}\,,
\]
where $f\equiv 1$ on the support of $X$. The triple $(\Lambda_{\F}\V(\Mcal),\{.,.\},\delta_S)$ is an algebraic structure called \textit{differential Gerstenhaber algebra}. In the quantized theory this structure is upgraded to a \textit{BV algebra} by introducing a certain degree 1 operator. In what follows we will 
use a slightly formal notation $X(\ph)=\int  X_x(\ph)\frac{\delta}{\delta \ph(x)}$ for a vector field $X\in\V(\Mcal)$. This is analogous to the notation $v=\sum_{i=1}^Nv^i\partial_i$ used for a vector field $v$ on an $N$-dimensional manifold. This notation allows to make contact with the standard physics literature on the BV formalism, if one identifies $\frac{\delta}{\delta \ph(x)}$ with a formal generator $\ph^\ddagger$, called \textit{antifield}.

The geometric structure described above appears also in theories where local symmetries are present, but there the space of multivector fields on an infinite dimensional manifold $\E(\Mcal)$ has to be replaced by the space of multivector fields on a certain \textit{graded}  infinite dimensional manifolds, which we denote by $\overline{\E}(\Mcal)$. We will show how this space is constructed on the example of Yang-Mills theories and the free electromagnetic field.

Let $G$ be a finite dimensional semisimple compact Lie group and $g\doteq Lie(G)$ its Lie algebra.We consider the trivial principal bundle $P=M\times G$ over $M$. In general one could also consider non-trivial bundles, but we take the point of view that the QFT model on a given spacetimes should be first constructed from ``simple building blocks'', i.e. algebras associated to regions that are topologically simple, and the global structure is recovered from the properties of the net itself. This suggests that also in case of gauge theories it makes sense to start with trivial bundles, since this is a good model for the local structure.

The configuration space $\E(\Mcal)$ for a Yang-Mills theory is the space of connection 1-forms, i.e. the space of $g$-valued 1-forms on $P$ which satisfy:
\begin{enumerate}
\item $R_{\alpha}^*\omega=\mathrm{ad}_{\al^{-1}}\circ \omega$,
\item $\omega(Z({\xi}))=\xi$,
\end{enumerate}
where $\al\in G$, $R_{\al}:G\rightarrow G$ is the right multiplication, $\mathrm{ad}$ is the adjoint representation of $G$ on $g$ and $Z:g\rightarrow \Gamma(TP)$ is the map which assigns to $\xi\in g$ its fundamental vector field given by
\[
Z_p(\xi)\doteq \frac{d}{dt}\Big|_{t=0}pe^{t\xi}\,.
\]
 $\E(\Mcal)$ can be equipped with an affine structure in the sense that the difference between two connection one forms $A$ and $A'$ is an element of the vector space $\Omega^1(M,g_P)$ of one forms on $M$ with values in the associated bundle $g_P\doteq P\times_Gg$. We will use this affine structure to define the notion of smoothness for the functionals on $\E(\Mcal)$ and to make $\E(\Mcal)$ into an infinite dimensional manifold, as we did for the scalar field. The derivative of a functional $F$ at $A_0\in\E(\Mcal)$ in the direction of $A\in\Omega^1(M,g_P)$ is defined as
\[
\left<F^{(1)}(A_0),A\right> \doteq \lim_{t\rightarrow 0}\frac{1}{t}\left(F(A_0 + tA) - F(A_0)\right)
\]
whenever the limit exists. The functional $F$ is called differentiable at $A_0$ if $F^{(1)}(A_0)$ exists for all $A \in \Omega^1(M,g_P)$. It is called continuously differentiable if it is differentiable at all points of $U$ and
$F^{(1)}:U\times X\rightarrow Y, (A_0,A)\mapsto \left<F^{(1)}(A_0),A\right>$
is a continuous map. It is called a $\Ccal^1$-map if it is continuous and continuously differentiable. Higher order derivatives are defined by the iteration of this definition. The infinite dimensional manifold structure on $\E(\Mcal)$ is induced by the topology $\tau_W$, defined in section \ref{genLagr}, with open neighborhoods given by
\[
W_{A_0}\doteq U+A_0=\{A_0+A|A\in U\}\,,
\]
where $U$ is an open neighborhood in $\E_c(\Mcal)\doteq\Omega_c^1(M,g_P)$, equipped with its standard inductive limit topology. Since we consider here only the trivial bundle $P=M\times G$, $\Omega^1(M,g_P)\cong \Omega^1(M,g)$ and $\E_c(\Mcal)\cong\Omega_c^1(M,g)$.

$\F(\Mcal)$ denotes again the space of multilocal compactly supported functionals on $\E(\Mcal)$ and $\V(\Mcal)$ is the space of multilocal vector fields. The generalized Lagrangian of the Yang-Mills theory is given by
\be\label{LagrYM}
L_M(f)(A)=-\frac{1}{2}\int_M f\,\tr(F \wedge * F)\,,
\ee
where $F=dA+\frac{1}{2}[A,A]$ is the curvature form corresponding to the connection 1-form $A$ and $*$ is the Hodge operator. The equation of motion reads:
\[
S'_{\Mcal}(A)=D_A\!*\!F=0\,,
\]
where $D_A$ is the covariant derivative induced by the connection $A$. To see that $H^1(\Lambda_{\F}\V(\Mcal),\delta_S)$ is non-trivial, we will construct explicitly non-trivial symmetries of the action corresponding to the Lagrangian \eqref{LagrYM}. Let us define the gauge group as the space of vertical $G$-equivariant compactly supported diffeomorphisms of $P$:
 \[
 \G:=\{\alpha\in \Diff_c(P)|\alpha(p\cdot g)=\alpha(p)\cdot g, \pi(\alpha(p))=\pi(p),\ \forall g\in G, p\in P\}\,.
 \]
We can also characterize $ \G$ as $\Gamma_c(M,P\times_G G)$ and for a trivial bundle $P$ this reduces ro $\Ci_c(M,G)$. It was shown (\cite{NeebWurz04,Gloe,Michor}, see also \cite{Neeb,Wock}) that $\Ci_c(M,G)$ can be equipped with the structure of an infinite dimensional Lie group modeled on its Lie algebra $\frakg_c(\Mcal):=\Ci_c(M,g)$. Since the gauge group is just a subgroup of $\Diff(P)$, it has a natural action on $\Omega^1(P,g)^G\cong \Omega^1(M,g)$ by the pullback. This induces the action of $\G(\Mcal)$ on $\E(\Mcal)$, and the corresponding derived action of $\frakg_c(\Mcal)$ is given by
\be\label{rho}
\rho_\Mcal(\xi)(A)=d\xi+[A,\xi]=D_A\xi\,.
\ee
The Yang-Mills action is invariant under the transformation \eqref{rho}, in the sense that
\[
\left<S'_{\Mcal}(A),\rho_\Mcal(\xi)(A)\right>=0\,,\quad\forall A\in\E(\Mcal)\,,
\]
so $\rho_M$ induces a map from $\frakg_c(\Mcal)$ to $\V(\Mcal)$, whose image is contained in the kernel of $\delta_S$. More generally, we can consider $\fG(\Mcal)\doteq\Ci_\ml(\E(\Mcal),\frakg_c(\Mcal))$, the space of multilocal functionals on the configuration space with values in the gauge algebra, and a map $\rho'_\Mcal:\fG(\Mcal)\rightarrow\V(\Mcal)$ defined by ${\rho'_\Mcal(\Xi)}(A)\doteq\rho_\Mcal(\Xi(A))A$, i.e.
\[
\partial_{\rho'_\Mcal(\Xi)}F(A)\doteq\left<F^{(1)}(A),\rho_\Mcal(\Xi(A))A\right>\,.
\]
Note that $\rho'$ is a natural transformation between $\fG$ and $\V$, both treated as functors from $\Loc$ to $\Vect$.

To see a more geometrical interpretation of the map $\rho'_\Mcal$ , note that $\fG(\Mcal)\subset\Gamma(\E(\Mcal)\times\frakg_c(\Mcal))$ (the space of sections of a trivial bundle over $\E(\Mcal)$), and we have a morphism of vector bundles $\E(\Mcal)\times\frakg_c(\Mcal)\rightarrow T\E(\Mcal)$ given by $(A,\xi)\mapsto (A,\rho_\Mcal(\xi) A)$. This actually equips  $\E(\Mcal)\times\frakg_c(\Mcal)$ with the structure of a Lie algebroid.

The invariance of the Yang-Mills action under $\rho_\Mcal$ implies that $\rho_\Mcal'(\fG(\Mcal))\subset \ker(\delta_S)$. In fact, one can characterize all non-trivial local symmetries this way, in the sense that for each $X\in\ker\delta_S$ there exists an element $\Xi\in\fG(\Mcal)$ and $I\in\delta_S(\La^2\V(\Mcal))$ such that
\[
X=I+\rho'_{\Mcal}(\Xi)\,.
\]
We can use this fact to kill the homology in degree one of the differential complex \eqref{K}. We extend the complex by adding  $\fG(\Mcal)$ in degree 2 and symmetric powers of $\fG(\Mcal)$ in higher degrees. We obtain a graded algebra $\KT(\Mcal)\doteq S^\bullet_{\F}\fG(\Mcal)\otimes_{\F}\La_{\F}\V(\Mcal)$ with a differential $\delta$ which acts on $\La_{\F}\V(\Mcal)$ as $\delta_S$ and on $S^\bullet_{\F}\fG(\Mcal)$ it is given by $\rho'_{\Mcal}$ extended by means of the graded Leibniz rule. The resulting differential complex is called the Koszul-Tate complex. On the functorial level it can be written as
 \be\label{KT}
\ldots\rightarrow\La^2\V\oplus\fG\xrightarrow{\delta=\delta_S\oplus\rho'}\V\xrightarrow{\delta=\delta_S}\F\rightarrow 0
\ee
The 0th homology of this complex is $\F_S(\Mcal)$ and higher homologies are trivial, so $(\KT,\delta)$ provides a resolution of $\F_S$.

We have already seen how to characterize the space of on-shell functionals in Yang-Mills theory, now we want to find a homological interpretation for the space of gauge invariant ones. This can be done with the use of the Chevalley-Eilenberg complex. The underlying algebra of this complex is \[\CE(\Mcal)\doteq\mathcal{O}_{\mathrm{ml}}(\overline{\mathfrak{E}}(\mathcal{M}))=\Ci_\ml(\E(\Mcal),\La\frakg'(\Mcal))\,,
\] the space of multilocal functionals on the graded manifold \[\overline{\E}(\Mcal)\doteq\E(\Mcal)\oplus\frakg(\Mcal)[1]\,,
\]
 where $\frakg(\Mcal):=\Ci(M,g)$.  The precise meaning of $\mathcal{O}_{\mathrm{ml}}$ is given in Refs. \cite{Book} and \cite{FR}. We call  $\overline{\E}(\Mcal)$ the \textit{extended configuration space}. The Chevalley-Eilenberg differential $\gamma$ is constructed in such a way that it encodes the action of the gauge algebra $\frakg(\Mcal) $ (note that we have dropped the restriction to compactly supported functions) on $\F(\Mcal)$. For $F\in\F(\Mcal)$ we define $\gamma F\in \Ci_\ml(\E(\Mcal),\frakg'(\Mcal))$ as
\be\label{ChE1}
(\gamma F)(A,\xi)\doteq (\rho_\Mcal(\xi)F)(A)\,,
\ee
where $\xi\in\frakg(\Mcal)$. For a form $\omega\in \frakg'(\Mcal)$, which doesn't depend on $A$ we set
\[
\gamma \omega(\xi_1,\xi_2)\doteq \omega([\xi_1,\xi_2])\,.
\]
Since $\gamma$ is required to be nilpotent of order 2 and has to satisfy the graded Leibniz rule, for a general $F\in\Ci_\ml(\E(\Mcal),\La^q\frakg'(\Mcal))$ we define
\begin{align*}
(\gamma F)(A,\xi_0,\dots,\xi_{q})\doteq& \sum_{i=0}^{q}(-1)^i\partial_{\rho_\Mcal(\xi_i)}(\iota_{(\xi_0,...,\hat\xi_i,...\xi_q)}F)(A)+\\
&+\sum_{i<j}(-1)^{i+j}F(A,[\xi_i,\xi_j],\ldots,\hat{\xi}_i,...,\hat{\xi}_j,...,\xi_q)\,,
\end{align*}
where the hat over a variable means that this variable is omitted and $\iota$ denotes the insertion of $n$ vector fields into an $n$-form. The differential complex looks as follows:
 \be\label{KT}
0\rightarrow\F(\Mcal)\xrightarrow{\gamma}\Ci_\ml(\E(\Mcal),\frakg'(\Mcal))\xrightarrow{\gamma}\Ci_\ml(\E(\Mcal),\La^2\frakg'(\Mcal))\rightarrow\ldots
\ee
Note that from \eqref{ChE1} it follows that the kernel of $\gamma$ in degree 0 consists of all the multilocal functionals invariant under $\rho'_\Mcal$. Hence $H^0(\CE(\Mcal),\gamma)=\F^{\inv}(\Mcal)$, the space of invariants. Since all the constructions here are functorial and maps between functors are natural transformations, we can drop the argument $\Mcal$, whenever it doesn't lead to confusion. 
\begin{rem}
Formally we can write elements of $\CE$ as sums of functionals of the form
\[
F(A)(\xi_1,\ldots,\xi_n)=\sum_{a_1,...,a_n}\int f(A)(x_1,...,x_n)_{a_1,...,a_n}\xi_1(x_1)^{a_1}\ldots \xi_n(X_n)^{a_n}d\mu_g(x_1)\ldots d\mu_g(x_n)\,,
\]
where $f(A)\in\Gamma'_{n}(M^n,g^{\otimes n})$ is an antisymmetric distributional section. Let us denote by $c$ the evaluation functional $c^a(x)(\xi)\doteq\xi^a(x)$. Clearly $c^a(x)\in\frakg'$. We call these evaluation functionals \textit{ghosts} and we write
\[
F(A)(\xi_1,\ldots,\xi_n)=\sum_{a_1<...<a_n}\int f(A)(x_1,...,x_n)_{a_1,...,a_n}c(x_1)^{a_1}\wedge\dots \wedge c(x_n)^{a_n}d\mu_g(x_1)\ldots d\mu_g(x_n)\,.
\]
\end{rem}

The Chevalley-Eilenberg complex and the Koszul-Tate complex fit together into one structure called the BV complex. To see how it arises in a natural way it is worth to look back at the example of the scalar fields, which we discussed at the beginning of this subsection. There, in order to characterize the space of on-shell functionals we needed to consider the space of multilocal vector fields on the configuration space. Now, to take the gauge symmetries into account, we need to extend the configuration space into a graded manifold $\overline{\E}$. The space of multivector fields on $\overline{\E}$ is given by
\[
\BV\doteq\mathcal{O}_{\mathrm{ml}}(T^*[-1]\overline{\mathfrak{E}})\,,
\]
the algebra of functions on \[
T^*[-1]\overline{\E}=\E\oplus\frakg[1]\oplus\E^*[-1]\oplus\frakg^*[-2]\,,
\] 
the odd cotangent bundle of $\overline{\E}$ , with the negative grading on the fiber.  Here $\mathfrak{E}^*\doteq \Gamma(E^*\rightarrow M)$, $\mathfrak{g}^*\doteq\mathcal{C}^{\infty}(M,g^*)$ and $E^*$, $g^*$ are algebraic duals. The numbers in square brackets indicate grading. The geometric interpretation which we present here fits very well with the spirit of the functional approach. Constructing the underlying algebra of the BV complex we are working all the time with multilocal functionals but we have to pass from infinite dimensional manifolds to graded infinite dimensional manifolds. Clearly, both $\CE$ and $\KT$ with inverted grading are subalgebras of $\BV$. We will now extend the differentials $\delta$ and $\gamma$ to the whole of $\BV$.

As in the case of the scalar field, $\BV$ can be equipped with the graded Schouten bracket $\{.,.\}$. We can use this structure to extend the Koszul-Tate differential to the whole algebra $\BV$ by setting
\[
\delta X\doteq \{X,S\}\,,\qquad X\in\BV\,.
\]
The Chevalley-Eilenberg differential can also be written in terms of the bracket in a similar manner.
We can find a natural transformation $\theta:\D\rightarrow\BV$ such that 
\[
\gamma X= \{X,\theta_\Mcal(f)\}\,,
\]
where $f\equiv 1$ on $\supp f$ and $X\in\CE$. The notion of natural Lagrangians introduced in section \ref{genLagr} applies to natural transformations from $\D$ to $\BV$ as well as it applied to those from $\D$ to $\F$, and we can treat $\theta$ as such a natural Lagrangian. This motivates the following notation;
\[
\gamma X\doteq \{X,\theta\}\,,\qquad X\in\BV\,.
\]
Next we define the BV differential as the sum
\[
s=\delta+\gamma=\{.,S+\theta\}\,,
\]
and call $S^\ex\doteq S+\theta$ the \textit{extended action} of the Yang-Mills theory.

The algebra $\BV(\Mcal)$ is equipped with two gradings. One is the grading of the Chevalley-Eilenberg complex, called the pure ghost number $\#\pg$, and the other is the antifield number $\#\af$ which is 0 for functions on $\overline{\E}(\Mcal)$, 1 for the elements of $\V(\Mcal)$ and 2 for the elements of $\fG(\Mcal)$. Using these two gradings we can construct a bicomplex
\[
\begin{CD}
\ldots@>\delta>>\big(\La^2\V\oplus\fG\big) @>\delta>>\V@>\delta>>\F@>\delta>>0\\ 
@.     @VV{\gamma}V@VV{\gamma}V@VV{\gamma}V@.\\
\ldots@>\delta>>{\Ci_\ml\big(\E,(\La^2\E_c\oplus\frakg_c)\widehat{\otimes}\frakg'}\big)@>\delta>>{\Ci_\ml\big(\E,\E_c\widehat{\otimes}\frakg'\big)}@>\delta>>{\Ci_\ml\big(\E,\frakg'\big)}@>\delta>>0
\end{CD}
\]
Note that the first row is just the Koszul-Tate complex. 
The total grading $\#\gh=\#\pg-\#\af$ is called the \textit{ghost number},  and a standard result in homological algebra tells us that the cohomology of the total complex is given by
\[
H^k(\BV(\Mcal),s)=H^k(H_0(\BV(\Mcal),\delta),\gamma)\,.
\]
Note that taking the 0th homology of $\delta$ amounts to going on-shell, while taking the 0th cohomology of $\gamma$ characterizes gauge invariants. Hence, 
\[
H^0(\BV(\Mcal),s)=\F_S^\inv(\Mcal)\,,
\]
so we obtain a homological interpretation of the space of gauge invariant on-shell functionals. Again we have a differential Gerstenhaber algebra $(\BV(\Mcal),\{.,.\},s)$. We can use this structure to implement the gauge fixing by modifying the extended action. Note that the $\#\af=0$ term of $S^\ex$ is still the original Yang-Mills action, which doesn't induce normally hyperbolic equations of motion, so we cannot construct retarded and advanced Green's operators. The idea now is to find an automorphism $\al$ of $(\BV(\Mcal),\{.,.\})$ such that $\tilde{S}\doteq \al(S^\ex)$ at  $\#\af=0$ induces normally hyperbolic equations of motion. A concrete form of $\al$ depends on the choice of gauge fixing and for particular choices one might need to extend $\BV(\Mcal)$ with some further generators (antighosts, $b$ fields). Let us define $\tilde{s}\doteq \al\circ s\circ \al^{-1}=\{.,\tilde{S}\}$. Obviously we have $H^0(\al(\BV(\Mcal)),\tilde{s})=\F_S^\inv(\Mcal)$, so we didn't loose the information about the gauge invariant on-shell observables. On the other hand, the  $\#\af=0$ term of the new extended action  $\tilde{S}$ induces normally hyperbolic equations of motion, so we can find $\Delta_{\tilde{S}}^A$ and $\Delta_{\tilde{S}}^R$ and introduce the Peierls bracket $\{.,.\}_{\tilde{S}}$ on $\BV(\Mcal)$. It turns out that $\tilde{s}$ is a derivation with respect to this bracket, so $\{.,.\}_{\tilde{S}}$ is well defined on $\F_S^\inv(\Mcal)$. This concludes the construction of the classical theory on $\Mcal$ in the presence of local symmetries.
\subsection{Example: Electromagnetic field}
Let us illustrate the general construction described above on the example of the electromagnetic field. The gauge group is $G=U(1)$, so $g=\RR$ and the Lagrangian takes the form
\[
L_{\Mcal}(f)(A)=-\frac{1}{2}\int_M f\,(F \wedge * F)\,.
\]
$\E(\Mcal)$ is the space of principal connections on $M\times U(1)$ and it is an affine space modeled on  $\E_c(\Mcal)=\Omega_c^1(M)$. As in the case of the free field we can consider the space $\F_{\mathrm{lin}}(\Mcal)$ of linear functionals on $\E(\Mcal)$. They are of the form
\[
F_{\beta}(A)=\int_M A\wedge *\beta\,,
\]
We can now apply to  $\F_{\mathrm{lin}}(\Mcal)$ the general BV formalism and compare the construction of \cite{Dim92}. The equation of motion is given by
\[
\delta dA=0\,,
\]
so the image of $\delta_S$ consists of functionals $F_{\beta}$, where $\beta =\delta d\eta$ for some $\eta\in\Omega_c^1(M)$. We can realize  $\F_{\mathrm{lin},S}(\Mcal)$ as the space of equivalence classes of forms 
\[
\F_{\mathrm{lin},S}(\Mcal)\cong\frac{\Omega_c^1(M)}{\delta d\Omega_c^1(M)}\,.
\]
Now we have to characterize the kernel of $\gamma$. It consists of linear functionals that satisfy
\[
0=(\gamma F_{\beta})(c)=\int_M dc\wedge *\beta=\int_M c\wedge *\delta\beta\,.
\]
It follows that $\delta \beta=0$. Let us denote $\Omega_{c,\delta}^1(M)\doteq \{\omega\in\Omega_c^1(M)|\delta\omega =0\}$. The space of gauge invariant on-shell linear functionals is isomorphic to (compare with \cite{SDH14,DS11,DL12})
\[
\F_{\mathrm{lin},S}^\inv(\Mcal)\cong \frac{\Omega_{c,\delta}^1(M)}{\delta d\Omega_c^1(M)}\,.
\]
Among these functionals we can distinguish the ones which are constructed from the field strength, i.e. those of the form
\[
\int_M dA\wedge *\eta=\int_M A\wedge *\delta\eta=F_{\delta\eta}(A)\,.
\]
If $H^1(M)$ is trivial, then all elements of $\F_{\mathrm{lin},S}^\inv$ arise from field strength functionals, since all co-closed forms are also co-exact.

\subsection{Quantization}
In this section we discuss quantization along the lines of \cite{FR3}. We start with the discussion of the free scalar field. We consider the deformation of $\delta_{S_0}$ under the time-ordering operator $\TT$. This deformation corresponds to the difference between the ideal generated by {\eom}'s in the classical theory (i.e. with respect to ``$\cdot$'') and the ideal generated by {\eom}'s with respect to $\T$. We define
\be
\delta_{S_0}^{\TT}\doteq\mathcal{T}^{-1}\circ\delta_{S_0}\circ\mathcal{T}\,,
\ee
Let us first consider regular functionals. Explicit computation shows that, on $\F_{\reg}(\Mcal)$, 
\[
\delta_{S_0}^{\TT}=\delta_{S_0}-i\hbar\Lap\,, 
\]
where $\Lap$ acts on vector fields $X\in\V_{\reg}(\Mcal)$ as
\[
\Lap X(\ph)=-\int \frac{\delta X_x}{\delta \ph(x)}(\ph),\qquad \textrm{where}\ X(\ph)=\int  X_x(\ph)\frac{\delta}{\delta \ph(x)}\,,
\]
It's remarkable that the operator $\Lap$ is ``almost'' a derivation of $\La\V_\reg(M)$ and the failure is characterized by $\{.,.\}$, i.e:
\[
\Lap(X\wedge Y)-(-1)^{|Y|}\Lap(X)\wedge Y-X\wedge \Lap (Y)=(-1)^{|Y|}\{X,Y\}\,,
\]
The triple $(\La\V_\reg(\Mcal),\{.,.\},\Lap)$ is an example of a mathematical structure called \textit{BV algebra}. Physically the relation between $\delta_{S_0}^{\TT}$ and $\delta_{S_0}$ corresponds to the \textit{Schwinger-Dyson equation}. Let $X(\ph)=\int  X_x(\ph)\frac{\delta}{\delta \ph(x)}$. We obtain
\[
-(\delta_{S_0}^{\TT}X)(\ph)=\TT^{-1}\int\!\! \left(\TT X_x\cdot\frac{\delta S_0(\ph)}{\delta \ph(x)}\right)(\ph)=\int\!\!   X_x(\ph)\frac{\delta S_0(\ph)}{\delta \ph(x)}-i\hbar\!\! \int\!\! \frac{\delta X_x}{\delta \ph(x)}(\ph)\,,
\]
where $\frac{\delta S_0(\ph)}{\delta \ph(x)}$ is a shorthand notation for $\frac{\delta L_0(f)(\ph)}{\delta \ph(x)}$, where we take the limit $f\rightarrow 1$. Note that 
\[
\int\!\!   \TT X_x(\ph)\frac{\delta S_0(\ph)}{\delta \ph(x)}=\int\!\!   \left(\TT X_x\star\frac{\delta S_0}{\delta \ph(x)}\right)(\ph)\,.
\]
Hence, 
\[
\TT\!\int \!\! \left(X_x\cdot\frac{\delta S_0(\ph)}{\delta \ph(x)}\right)(\ph)=i\hbar\, \TT\!\int \frac{\delta X_x}{\delta \ph(x)}(\ph)\,,\] 
modulo the $\star$-ideal generated by the {\eom}'s, which is exactly the algebraic Schwinger-Dyson equation. For gauge theories, one simply replaces $\La_{\F}
\V_\reg(\Mcal)$ with $\BV_{\reg}(\Mcal)$ and $S_0$ is the $\#\af=0$ quadratic term of the extended action.

Let us now consider a deformation of $\delta_{S_0}$ which corresponds to introducting the interaction. Here we treat the scalar field and gauge theories together, but still restrict to regular functionals, i.e. $V\in\BV_{\reg}(\Mcal)$  is a regular interaction term.  We define the \textit{quantum BV operator} $\hat{s}$ as
\be\label{intertwining:s}
\hat{s}\doteq R_{V}^{-1}\circ\delta_{S_0}\circ R_{V}\,.
\ee
Let us assume that
\be\label{qme}
\delta_{S_0}\big(e_{\sst\TT}^{i V/\hbar}\big)=0\,.
\ee
This condition reduces to the known \textit{quantum master equation} {\qme}, since
\be\label{qmenr}
\delta_{S_0}\big(e_{\sst\TT}^{i V/\hbar}\big)=\frac{1}{2}\{S_0+V,S_0+V\}-i\hbar\Lap V\,,
\ee
where we set $\{S_0,S_0\}\equiv 0$, since $S_0$ doesn't contain antifields. If \eqref{qme} holds, then
\be\label{snr}
\hat{s}F=\{F,S_0+V\}-i\hbar \Lap F\,,
\ee
for $F\in \F_{\reg}(\Mcal)$. If {\qme} is fulfilled, then the cohomology of $\hat{s}$ characterizes the space of \textit{quantum} gauge invariant on-shell observables.

Now we want to extend the {\qme} and $\hat{s}$ to local functionals. This is done by renormalizing the time-ordered products present in \eqref{intertwining:s} and the \eqref{qme} using the Epstein-Glaser framework. Clearly, formulas \eqref{qmenr} and \eqref{snr} are not well defined for local arguments, since $\Lap$ is singular. Nevertheless, very similar results can be obtained using the anomalous Master Ward Identity (\cite{BreDue,H}), which states that there exists a family of maps 
\be
\widetilde{\Lap}^n:\TT(\BV_\loc(\Mcal))^{n
+1}\to \fA_\loc(\Mcal)\,,
\ee
which depend locally on their arguments and the formal power series 
\[
\widetilde{\Lap}(V)\doteq\sum_{n=0}^{\infty}\widetilde{\Lap}^n(V^{\otimes n};V)\,,
\]
fulfills the identity
\be\label{MWI}
\int \left(e_{\sst{\TT}}^{iV/\hbar}\T \frac{\delta V}{\delta \ph(x)}\right)\star\frac{\delta S_0}{\delta\ph(x)}=e_{\sst{\TT}}^{iV/\hbar}\T(\tfrac{1}{2}\{V+S_0,V+S_0\}_{\TT}-\widetilde{\Lap}(V))\,,
\ee
The maps $\widetilde{\Lap}^n$ can be determined recursively. For an explicit formula, see \cite{BreDue,Rej13}. We can now see that the renormalized {\qme} reduces to 
\be\label{qmer}
\delta_{S_0}\big(e_{\sst\TT}^{i V/\hbar}\big)=\frac{1}{2}\{S_0+V,S_0+V\}-i\hbar\widetilde{\Lap}(V)\,.
\ee
and the renormalized quantum BV operator takes the form
\be\label{sr}
\hat{s}F=\{F,S_0+V\}-i\hbar \Lap_VF\,,
\ee
where $\Lap_V(F)\doteq\frac{d}{d\la}\Big|_{\la=0}\widetilde{\Lap}(V+\la F)$. Note that the renormalized operator $\Lap_V$ depends on $V$, in contrast to the non-renormalized one $\Lap$. It has no longer the interpretation of a graded Laplacian, but is still a functional differential operator.

\subsection{Remarks on topological obstructions}\label{topo}
Let us now briefly discuss the possibility to extend the QFT functor to topologically non-trivial spacetimes. On physical grounds, we expect that theories like QED should be sensitive to the topology of the underlying spacetime (specifically to $H^2(M)$). This could potentially produce a topological obstruction to extending the QFT functor. Indeed, already on Minkowski spacetime one can see this kind of mechanism in the violation of Haag duality for non-contractible regions for QED \cite{F82}. The topological nature of this mechanicsm can be best seen with the use of the local cohomology framework due to Roberts \cite{Roberts76,Roberts77}. Later on this line of research was continued in \cite{RR06,CRV}. A different, but related homological analysis of the problem was recently performed in \cite{BSS14} (see also \cite{BDHS13,BDS14} for earlier work on the subject). In the BV framework outlined in the present section, such global effects can be seen in the fact that the $\BV$ functor (as a functor into topological differential graded algebras) fails to be injective if extended to topologically non-trivial spacetimes. 
\appendix
\section{Mathematical structures}
\subsection{Categories and functors}\label{categories}
Here we recall some basic notions of category theory, which are used in locally covariant quantum field theory.
\begin{df}
A category $\Ca$ consists of:
\begin{itemize}
\item a class of objects $\obj(\Ca)$,
\item a class of morphisms (arrows) $\hom(\Ca)$, such that each $f\in\hom(\Ca)$ has a unique \textbf{source object} and  \textbf{target object} (both are elements of $\obj(\Ca)$). For a fixed $a,b\in\obj(\Ca)$, we denote by $\hom(a,b)$ the set of morphisms with $a$ as a source and $b$ as a target,
\item a binary associative operation $\circ:\hom(a,b)\times\hom(b,c)\rightarrow \hom(a,c)$, $f,g\mapsto f\circ g$, called composition of morphisms,
\item the identity morphism $\id_c$ for each $c\in\obj(\Ca)$.
\end{itemize}
\end{df}
\begin{df}
Let $\Ca$, $\Da$ be categories. A covariant functor $F$ assigns to each object $c\in\Ca$ an object $F(c)$ of $\Da$ and to each morphism $f\in\hom(\Ca)$, a morphism $F(f)\in\hom(\Da)$ in such a way that the following two conditions hold:
\begin{itemize}
    \item${F}(\mathrm{id}_{c}) = \mathrm{id}_{{F}(c)}$ for every object $c \in \Ca$.
    \item${F}(g \circ f) = {F}(g) \circ {F}(f)$ for all morphisms $f:a \rightarrow b$ and $g:b\rightarrow c.\,\!$
\end{itemize}
\end{df}
Next we recall the definition of a tensor category, which is important in the context of Einstein causality of the QFT functor.
\begin{df}
We call a category $\mathbf{C}$ strictly monoidal (tensor category) if there exists a bifunctor $\otimes:\mathbf{C}\times\mathbf{C}\rightarrow\mathbf{C}$ which is associative, i.e. $\otimes(\otimes\times 1)=\otimes( 1\times\otimes)$ and there exists an object $e$ which is a left and right unit for $\otimes$.
\end{df}
\bibliographystyle{amsplain}
\subsection{Infinite dimensional differential geometry}\label{smooth}
Let us start with the definition of smoothness for functionals on general locally convex topological vector spaces.
\begin{df}[after \cite{Neeb}]\label{smooth0} 
Let $X$ and $Y$ be  locally convex topological vector spaces, $U \subseteq X$ an open set and $f:U \rightarrow Y$ a map. The derivative of $f$ at $x$ in the direction\index{derivative!on a locally convex vector space} of $h$ is defined as
\be\label{de}
f^{(1)}(x)(h) \doteq \lim_{t\rightarrow 0}\frac{1}{t}\left(f(x + th) - f(x)\right)
\ee
whenever the limit exists. The function $f$ is called differentiable\index{infinite dimensional!calculus} at $x$ if $f^{(1)}(x)(h)$ exists for all $h \in X$. It is called continuously differentiable if it is differentiable at all points of $U$ and
$df:U\times X\rightarrow Y, (x,h)\mapsto df(x)(h)$
is a continuous map. It is called a $\Ccal^1$-map if it is continuous and continuously differentiable. Higher derivatives are defined for $\Ccal^n$-maps by 
\be
f^{(n)} (x)(h_1 , \ldots , h_n ) \doteq \lim_{t\rightarrow 0}\frac{1}{t}\big( f^{(n-1)} (x + th_n )(h_1 , \ldots, h_{n-1} ) -
 f^{(n-1)} (x)(h_1 , \ldots, h_{n-1}) \big)
 \ee
\end{df}
As a consequence of this definition, if $F$ is a smooth functional on $\E(\Mcal)$, then its $n$-th derivative at the point $\ph\in\E(\Mcal)$ is a compactly supported distributional density $F^{(n)}(\ph)\in\Ecal'(M^n)$, where $\mathcal{E}(M^n)$ is $\mathcal{C}^{\infty}(M^n,\mathbb{R})$ with its standard Fr{\'e}chet topology. We can use the distinguished volume form $d\mu_g$ on $M$ to construct densities from functions and to provide an embedding of $\Dcal(M^n)$, the space of compactly supported functions on $M^n$, 
 into $\Ecal'(M^n)$. For more details on distributions on manifolds, see chapter 1 of \cite{Baer}. Using the distinguished volume form we can identify derivatives $F^{(n)}(\ph)$ with distributions.
 
 A natural question to be asked at the point is how the notion of smoothness introduced by definition \ref{smooth0} relates to the smooth structure induced by the topology $\tau_W$ (i.e. how $F^{(1)}$ relates to $DF$). In fact, it was shown in \cite{BFR} that these two notions are equivalent. The argument goes as follows: let $F$ be a smooth compactly supported functional with support $K\subset M$; consider a map $i_\chi:\E(\Mcal)\rightarrow\ph_0+\D(\Mcal)$ defined by
\[
i_\chi(\ph)\doteq\ph_0+(\ph-\ph_0)\chi\,,
\]
where $\chi\in\D(\Mcal)$ satisfies $\chi\equiv 1$ on $K$. The map $i_\chi$ is smooth from $\E(\Mcal)$ endowed with the compact open topology to $U\subset\E(\Mcal)$ endowed with $\tau_W$. Here $U$ is the connected, 	$\tau_W$-open neighborhood of $\ph_0$ so $i_\chi^{-1}(U)$ is open in the compact-open topology and
\[
\left<F^{(1)}(\ph),\vec{\ph}\right>=\left<F^{(1)}(\ph),\chi\vec{\ph}\right>=\left<DF(\ph),\chi\vec{\ph}\right>\,,
\]
where $\vec{\ph}\in i_\chi^{-1}(U)$. This generalizes also to higher derivatives. Note that the crucial property which allows us to switch between the compact-open and the $\tau_W$ topology is the fact that the functionals we are dealing with are compactly supported.
\subsection{H\"ormander topology $\tau_\Xi$}\label{Ht}
Let $\Ecal'_{C}(M^n)$ denote the space of compactly supported distributions on $M^n$, with WF sets contained in a conical set $C\subset T^*M^n$. Now let  $C_n\subset \Xi_n$ be a closed cone contained in $\Xi_n$ defined by \eqref{cone}. We introduce (after \cite{Hoer1,BaerF,BDF}) the following family of seminorms on $\Ecal'_{C_n}(M^n)$: 
\[
p_{n,\ph,\tilde{C},k} (u) = \sup_{{k}\in \tilde{C}}\{(1 + |{k}|)^k |\widehat{\ph u}({k})|\}\,,
\]
where the index set consists of $(n,\ph,\tilde{C},k)$ such that $k\in \NN_0$, $\ph\in \Dcal(M)$ and $\tilde{C}$ is a closed cone in $\RR^n$ with $(\supp ( \ph ) \times \tilde{C}) \cap C_n = \varnothing$. These seminorms, together with the seminorms of the weak topology provide a defining system for a locally convex topology denoted by $\tau_{C_n}$. To control the wave front set properties inside open cones, we take an inductive limit.  The resulting topology is denoted by $\tau_{\Xi_n}$. For microcausal functionals $F^{(n)}(\ph)\in\Ecal'_{\Xi_n}(M^n)$, so we can equip $\F_\mc(\Mcal)$ with the initial topopolgy with respect to mappings:
\be\label{tauH}
\Ci(\E(\Mcal),\RR)\ni F\mapsto F^{(n)}(\ph)\in(\Ecal_{\Xi_n}(M^n),\tau_{\Xi_n})\quad n\geq0\,,
\ee
This topology is denoted by $\tau_\Xi$.
\bibliography{References}
 \end{fmffile}
\end{document}